\documentclass[12pt]{iopart}
\usepackage{iopams}
\usepackage{graphicx}
\usepackage{bm}
\usepackage{dcolumn}
\usepackage{color}
\usepackage[latin9]{inputenc}
\usepackage{url}
\usepackage{subfigure}
\usepackage{float}
\usepackage{longtable}
\usepackage{epstopdf}

\sloppypar

\newcommand{\raiseentry}[1]{\smash{\raise 0.7 em \hbox{#1}}}

\newenvironment{equationarray*}
{\arraycolsep 0.14 em
\begin{eqnarray*}}
{\end{eqnarray*}}

\begin{document}

\title{Classification methods for noise transients in advanced gravitational-wave detectors}

\author{Jade Powell$^1$, Daniele Trifir\`o$^2$ $^5$, Elena Cuoco$^3$ $^4$, Ik Siong Heng$^1$ and Marco Cavagli\`a$^5$}

\address{$^1$ SUPA, Institute for Gravitational Research, School of Physics and Astronomy, University of Glasgow, Glasgow, G12 8QQ, Scotland, United Kingdom.}
\address{$^2$ Dipartimento di Fisica E. Fermi, Universit\`a di Pisa, Pisa 56127, Italy.}
\address{$^3$ European Gravitational Observatory (EGO), Via E Amaldi, I-56021 Cascina, Italy}
\address{$^4$ Istituto Nazionale di Fisica Nucleare (INFN) Sez. Pisa Edificio C - Largo B. Pontecorvo 3, 56127 Pisa, Italy.}
\address{$^5$ Department of Physics and Astronomy, The University of Mississippi, University, MS 38677, USA.}

\date{\today}

%%%%%%%%%%%%%%%%%%%%%%%%%%%%%%%%%%%%%%%%%%%%%%%%%%%%%%%%%%%%%%%%%%%%%%%%%%%%%%%
%%%%%%%%%%%%%%%%%%%%%%%%%%%%%%%%%%%%%%%%%%%%%%%%%%%%%%%%%%%%%%%%%%%%%%%%%%%%%%%
%%%%%%%%%%%%%%%%%%%%%%%%%%%%%%%%%%%%%%%%%%%%%%%%%%%%%%%%%%%%%%%%%%%%%%%%%%%%%%%

\begin{abstract}
Noise of non-astrophysical origin will contaminate science data taken by the 
Advanced Laser Interferometer Gravitational-wave Observatory (aLIGO) and 
Advanced Virgo gravitational-wave detectors. Prompt characterization of 
instrumental and environmental noise transients will be critical for improving 
the sensitivity of the advanced detectors in the upcoming science runs. 
During the science runs of the initial gravitational-wave detectors, noise 
transients were manually classified by visually examining the time-frequency 
scan of each event. Here, we present three new algorithms designed for the 
automatic classification of noise transients in advanced detectors. Two of 
these algorithms are based on Principal Component Analysis. They are Principal 
Component Analysis for Transients (PCAT), and an adaptation of LALInference 
Burst (LIB). The third algorithm is a combination of an event generator called 
Wavelet Detection Filter (WDF) and machine learning techniques for 
classification. We test these algorithms on simulated data sets, and we show their 
ability to automatically classify transients by frequency, SNR and waveform 
morphology.

\end{abstract}

%%%%%%%%%%%%%%%%%%%%%%%%%%%%%%%%%%%%%%%%%%%%%%%%%%%%%%%%%%%%%%%%%%%%%%%%%%%%%%%
%%%%%%%%%%%%%%%%%%%%%%%%%%%%%%%%%%%%%%%%%%%%%%%%%%%%%%%%%%%%%%%%%%%%%%%%%%%%%%%
%%%%%%%%%%%%%%%%%%%%%%%%%%%%%%%%%%%%%%%%%%%%%%%%%%%%%%%%%%%%%%%%%%%%%%%%%%%%%%%

\section{Introduction}
\label{section:introduction}

The sensitivity of advanced gravitational-wave detectors will be limited by 
multiple sources of noise from the hardware subsystems and the environment. The 
Advanced Laser Interferometer Gravitational-wave Observatory (aLIGO) detectors 
\cite{2010CQGra..27h4006H, 0264-9381-32-7-074001}, which are expected to become fully operational in late summer 2015, and 
Advanced Virgo \cite{2008CQGra..25k4045A}, which is expected to become fully operational in 2016, will 
include upgrades to all hardware subsystems including suspensions, lasers, seismic isolation and optics. The upgrades are designed to 
reduce noise sources and significantly improve the sensitivity of the initial LIGO and Virgo instruments.
 
The low frequency sensitivity of the detectors ($\lesssim 10$ Hz) will be limited by the effects of seismic 
noise. Thermal noise due to Brownian motion will 
be the most dominant noise source in the most sensitive frequency range of the instruments. At 
frequencies higher than $\sim150$ Hz, shot noise, due to quantum uncertainties in the laser light, is 
expected to be the dominant noise source. Instrumental and environmental disturbances can also produce non-astrophysical triggers in science 
data, so called ``glitches," as well as increasing the false alarm rate of searches and producing a decrease in the
detectors' duty cycles. Although much has been learned from the initial LIGO phase, the advanced
detectors will be for the most part newly-designed instruments, never assembled or tested before.
Thus the success of the advanced detectors will require a huge effort in commissioning and detector characterization \cite{Christensen:2010zz, 2012CQGra..29o5002A}.
Understanding detector noise, which may affect the discovery of gravitational waves, will be critical for
increasing the chances of detecting an astrophysical gravitational-wave signal. 

aLIGO and Advanced Virgo are designed to perform searches for gravitational waves of various astrophysical 
origin \cite{0264-9381-32-7-074001}. Potential sources can be split roughly into two groups: (1) signals
with a known gravitational waveform, such as Compact Binary Coalescing (CBC) sources \cite{2007arXiv0710.1338P}, and (2) un-modelled 
sources, where the astrophysical source and the gravitational waveform may be completely unknown \cite{2012PhRvD..85l2007A}.  
 
Current astrophysical estimates on the rates of cosmic events detectable by advanced 
detectors \cite{2010CQGra..27q3001A} indicate that an advanced detector network will lead to multiple 
detections of gravitational-wave signals from CBC sources over the network operating time.  
An exciting new observational window on the Universe is within reach. The non-Gaussian and non-stationary nature of advanced detector noise may
produce glitches, which could affect the sensitivity of searches and be mistaken as
gravitational-wave detections, in particular for un-modelled
sources. To prevent this, searches for un-modelled gravitational-wave signals,
or ``bursts," currently combine the noise from multiple detectors coherently to
prevent a glitch being misinterpreted as an astrophysical signal \cite{2012PhRvD..85l2007A}. 

However, a glitch occurring at the same time in multiple detectors
could still lead to a false positive. If the origin of the noise cannot be identified and hardware improvements cannot be  made to remove the glitch, Data Quality (DQ) flags are applied.

DQ flags and ``vetoes" can be used to remove detector data that contains a high number of glitches
 \cite{2014arXiv1410.7764T}. This method requires monitoring multiple auxiliary channels, which are not sensitive to gravitational
waves, but provide important information about the environment and other degrees of freedom in the detector. 
Periods with a large number of glitches are flagged as likely to cause an adverse affect in the gravitational-wave data channel. 
DQ flags were used in the initial detector science runs, and were highly effective in increasing the sensitivity of searches \cite{2014arXiv1410.7764T}. The use of
DQ flags in Virgo's second science run gave an $\sim30\%$ increase in the volume of which Virgo is sensitive to CBC sources \cite{2012CQGra..29o5002A},
and $\sim5$ Mpc increase in the detection range of a binary neutron star system, with Signal to Noise Ratio (SNR) equal to 8, in the initial LIGO detectors \cite{2014arXiv1410.7764T}. DQ flags can lead to a lower false alarm rate, however, overzealous use of DQ flags can reduce the duty cycle of the instruments.

Glitch classification and categorization may provide valuable clues for
identifying the source of noise transients, and possibly lead to their
elimination. In previous LIGO and Virgo science runs, this classification was performed by visual inspection of the glitches' time series and/or spectrograms.
Visual inspection of individual glitches proved to be a slow and inefficient
method in attempts to categorize noise transients during the S6 LIGO science run. With even
more data expected in the advanced detector observing runs, faster and more
reliable techniques for the classification of noise transients are needed. In
order to increase detector sensitivity and duty cycle, it will be essential to
provide DQ information in real time during observing runs. This can
only be achieved with automatic glitch classification algorithms running in
low-latency as data is collected \cite{1742-6596-243-1-012006}.

We have developed three methods that can be used for the fast classification of advanced detector noise transients. 
They are Principal Component Analysis for Transients (PCAT), an adaptation of LALInference Burst (LIB) \cite{2014arXiv1409.2435E}, 
and a combination of a trigger generator called Wavelet Detection
Filter and Machine Learning techniques (WDF-ML) \cite{WDF-GRB} \cite{0264-9381-26-20-204022}. The three different methods are described in Section \ref{algorithms}. 
To test the performance of the classifying algorithms, we created three different simulated data sets, described in Section \ref{dataset}. 
These data sets are specifically designed to test the efficiency of the algorithms in classifying noise transients with 
different waveform morphology or frequency content. In Section \ref{Results} 
we present the results for the simulated data sets. This is followed by 
a discussion in Section \ref{discussion} of plans for future improvements, and a discussion of how our results 
may affect classification of real noise transients during the advanced detector science runs.

%%%%%%%%%%%%%%%%%%%%%%%%%%%%%%%%%%%%%%%%%%%%%%%%%%%%%%%%%%%%%%%%%%%%%%%%%%%%%%%
%%%%%%%%%%%%%%%%%%%%%%%%%%%%%%%%%%%%%%%%%%%%%%%%%%%%%%%%%%%%%%%%%%%%%%%%%%%%%%%
%%%%%%%%%%%%%%%%%%%%%%%%%%%%%%%%%%%%%%%%%%%%%%%%%%%%%%%%%%%%%%%%%%%%%%%%%%%%%%%

\section{Transient classifying algorithms}
\label{algorithms}

\begin{table}[t]
\footnotesize
\centering
\begin{tabular}{|p{0.1\textwidth}|p{0.25\textwidth}|p{0.25\textwidth}|p{0.25\textwidth}|}  
\hline
& PCAT & LIB & WDF-ML \\ \hline
  Pre-processing & Down-sample, high pass filter and whiten data. & Down-sample data. & Whiten data and apply a windowing procedure. Apply a wavelet transform to the data. \\ \hline
  Trigger detection & Triggers are amplitudes above a threshold on the standard deviation of the analyzed segment. & Does not generate its own triggers. For this study GPS times from injection log were used. & The triggers are the sum squared wavelet coefficients above a threshold value that are larger than the SNR.  \\ \hline
  Glitch Classification & Apply PCA to all glitches found in the data. Apply machine learning to PCA coefficients to classify glitches. & Create a signal model from a linear combination of PCs made from known glitch types. Apply nested sampling to h(t) data to calculate Bayes factors.  & Apply PCA and spectral embedding to reduce the data set. Apply machine learning to the reduced data for classification. \\ 
\hline
\end{tabular}
\caption{A summary of the three different glitch classifying algorithms used in this study.}
\label{codes}
\end{table}

\subsection{Principal Component Analysis for Transients}
PCAT is a python-based algorithm based on the use of Principal Component Analysis (PCA) 
\cite{jackson2003user} to identify and classify noise transients in LIGO channels.
PCA consists of a linear orthogonal transformation of a set of (possibly correlated) variables into a set of linearly uncorrelated variables, called Principal Components (PCs). The PCs define the direction
of greatest variation in the data. PCA allows a quick characterization of the intrinsic properties
of a data sample. It is a versatile and powerful method with a long history of applications in many
different fields \cite{Calder20011179,Bishop:2006:PRM:1162264,Lewicki:1998ea}.

A summary of the PCAT algorithm is given in Table 1. In this investigation, we use time-sampled values of LIGO's simulated h(t) strain as PCA input variables. The PCs are used to analyze the time variability of the channel and reconstruct the properties of the transients. While this article concentrates on noise transients in the time-domain,
PCAT also implements a frequency-domain analysis, where input variables
are Power Spectral Densities (PSDs). In general, PCA can be applied to any
set of observations with a number of variables.

\subsubsection{Pre-processing.}

The raw time series (sampled at $16384$ Hz) is first split into $32$ second-long segments with a 50\% overlap, then downsampled to $8192$ Hz, and high passed with a Butterworth 4th order filter with a $40$ Hz cutoff frequency. The data is then whitened by multiplying the Fast Fourier Transform (FFT) of the time series by the inverse of the square root of the detector's noise PSD, which is computed using the median-mean average algorithm, as described in \cite{Allen:2005cz}. The whitened FFT is inverted, yielding the whitened time series, of which the first and last $8$ seconds are discarded to avoid FFT artifacts at the edges. Noise transients are identified when the channel
amplitude exceeds a chosen threshold in units of the standard deviation of
the analyzed $16$ second segment. A value between $4.5$ and $5$ has been shown to maximize the algorithm efficiency in identifying transients, while minimizing false positives. For each set of points above the threshold (triggers), the time series is sampled with a fixed-width interval around the trigger's maximum amplitude (typically corresponding to around $125$ ms), and then rescaled to a maximum (absolute)
amplitude equal to one. This step is required to properly
compare the time series and identify the main features of different
transient families

\subsubsection{Basics of Principal Component Analysis.}
\label{basics}

The identified transients $n$ are used to construct an $m\times n$ data matrix ${\bf D}$, where the rows of the matrix 
are the time series of the transients of length $m$. The matrix is then standardized by subtracting the mean of each column from the data.
The standardized $m\times n$ data matrix ${\bf D_{s}}$ can be factored so that

\begin{equation}
	{\bf D_{s}} = {\bf U} \Sigma {\bf V}^{{\bf T}},
\label{standardizeX}
\end{equation}
where ${\bf U}$ is an $m \times m$ matrix with columns given by the eigenvectors of ${\bf D}_{s}^{T}{\bf D}_{s}$, ${\bf V}$ is an $n \times n$ matrix
with the eigenvectors of ${\bf D}_{s}{\bf D}_{s}^{T}$ as columns, and ${\bf \Sigma}$ is an $m\times n$ diagonal matrix. The rows of the matrix ${\bf U}$
are the PCs, which are ordered by decreasing eigenvalue absolute value. The diagonal values of ${\bf \Sigma}$ are
the eigenvalues of the PCs.
The data matrix ${\bf D_{s}}$ can be projected on the PC basis

\begin{equation}
	{\bf S}={\bf D_{s}}\ {\bf U}\,\label{scorematrix}.
\end{equation}
The $m\times n$ matrix ${\bf S}$ is called the Coefficient Matrix. The coefficients of the
expansion of the original data set w.r.t.\ the new basis are called PC coefficients. Transients with different features are expected to have different PC 
coefficients. These coefficients can be used to characterize the features of the 
transients. Since the PC eigenvectors are ordered by decreasing eigenvalues,
the first few coefficients typically identify the most important features of the 
glitch waveforms that can be separated from the noise. Glitch waveforms can be accurately
reconstructed from a linear combination of the first $k$ PCs, weighted by their respective coefficients,
where usually $k\ll n$.

The explained variance of the data is defined as

\begin{equation}
v(k) =  \frac{1}{\Sigma}\,\sum_{i=1}^{k}\Sigma_i\,,~~~~~\Sigma = \sum_{i=1}^n \Sigma_{i}\,,
\label{explvar}
\end{equation}
where $\Sigma_i$ are the eigenvalues of the matrix ${\bf \Sigma}$. The explained variance $0\le
v\le 1$ measures the variation (dispersion) of the data set as a function of its
dimensionality. The number of PCs that are needed to describe the sample up
to a given accuracy can be determined by setting a threshold in $v$. Thus PCA allows
dimensional reduction of the data set. An example of a PCAT variance curve is given in Figure 1.

\subsubsection{Classification.}
\label{sec:PCAT-classification}

\begin{figure}[!t]
\begin{centering}
        \includegraphics[width=7.5cm]{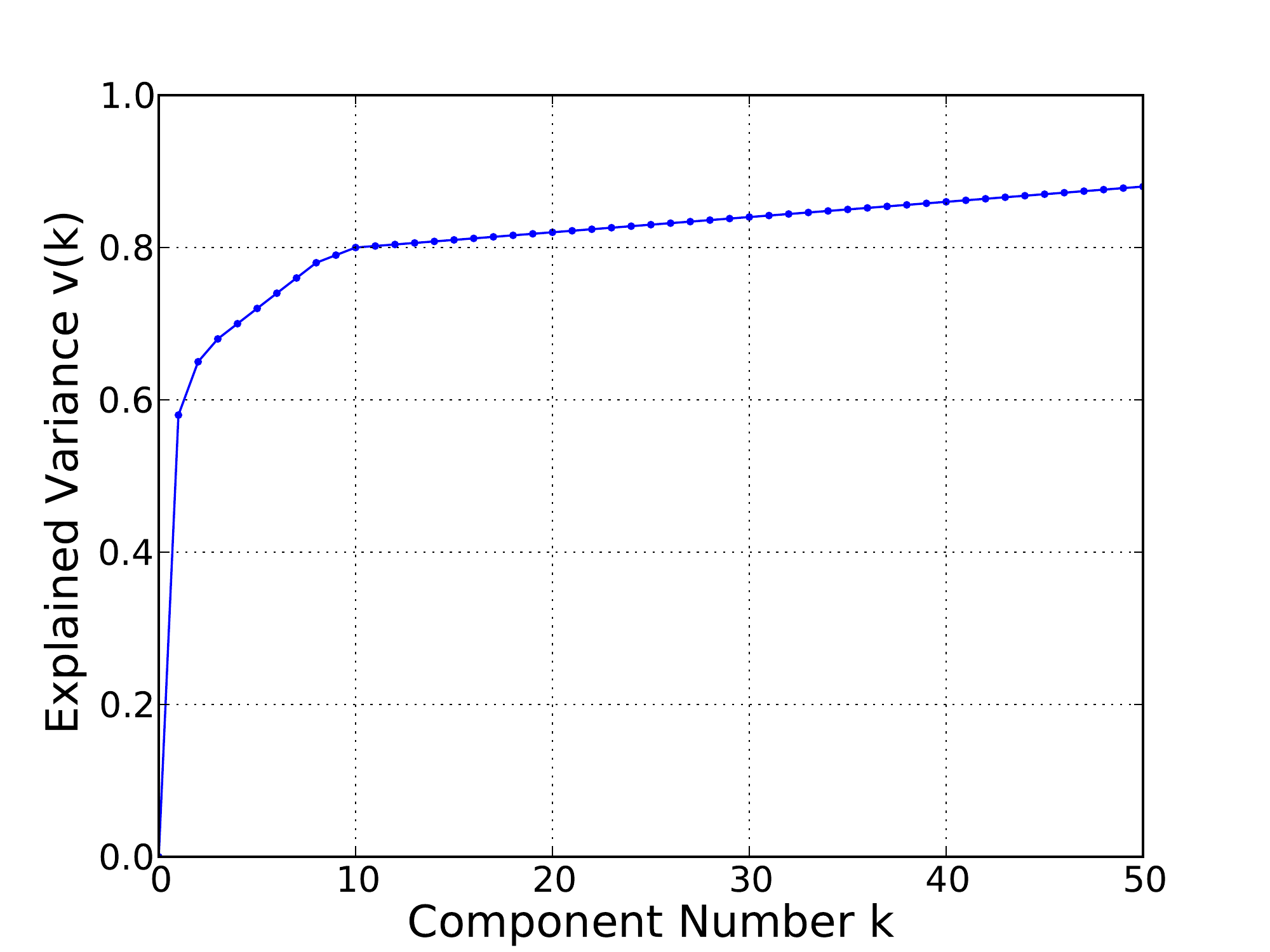}
	\label{fig:PCAT-MDC3-frequency-distribution}
\caption{PCAT explained variance as a function of the number of PCs, data set $1$. Changes in the variance curve help to find the ideal number of PCs.} 
\end{centering}
\end{figure}

PCAT uses the scikit-learn Gaussian Mixture Model (GMM) algorithm to 
cluster the PCA-reduced data \cite{scikit-learn}. The data are fit to a linear combination of
multivariate Gaussian distributions. The number of these distributions 
(number of classes) is determined by minimizing the Bayesian Information 
Criterion (BIC) \cite{Bhat:2010}

\begin{equation}
	BIC = - 2 \cdot \ln \mathcal{L} + k\cdot \ln(m)\,,
\label{biceq}
\end{equation}
where $m$ is the number of observations, $k$ is the number of free parameters to be estimated, and
$\mathcal{L}$ is the maximized value of the likelihood function. $\mathcal{L}$ depends on the
parameters of the model. An important feature of the $BIC$ algorithm is the calculation of a
penalty score for each of the free parameters in the data set to avoid over-fitting.  

Accurate classification of noise transients requires a careful choice 
of the number of PCs. A low number 
of PCs typically results in insufficient information to characterize the 
data. A high number of PCs leads to the inclusion 
of Gaussian noise features in the reduced dataset, which results in poor performance of the 
clustering algorithm. While no optimal method exists for choosing the ideal number 
of PCs, two of the most commonly used methods consist in setting a 
threshold on $v$, or looking for a slope change in the explained variance curve \cite{PCA-clustering}.

%%%%%%%%%%%%%%%%%%%%%%%%%%%%%%%%%%%%%%%%%%%%%%%%%%%%%%%%%%%%%%%%%%%%%%%%%%%%%%%%%%%%%%%%%%%%%%%%%%%%%%%%%%%

\subsection{LALInference Burst}

LALInference Burst (LIB) is a Bayesian parameter estimation algorithm for parameter estimation or model selection for gravitational-wave burst signals \cite{2014arXiv1409.2435E}. LIB is the burst adaptation of the LALInference library, which is designed for parameter estimation for CBC signals \cite{2014arXiv1409.7215V}. LIB uses nested sampling to calculate the Bayesian evidence with a sine Gaussian signal model \cite{Sivia1996, 2014arXiv1409.7215V}. 
It can produce posterior distributions for the parameters of the signal, such as the sky location \cite{2014arXiv1409.2435E}.

To adapt LIB for the classification of glitches, we adopt the PCA approach taken by Logue {\it et.\ al.} \cite{2012PhRvD..86d4023L, 2009CQGra..26j5005H} in 
their analysis of the explosion mechanism of core collapse supernovae. We take the time series of fifty glitches of a known type, sampled at 4096 Hz, and apply a second order 
Butterworth high pass filter at 40 Hz. We then FFT the waveforms, as LIB performs model selection in the frequency domain. PCA is then applied to the transient waveforms using the method described in Section \ref{basics}. A linear combination of 
the PCs, multiplied by the PC coefficients, is then used as the new signal model in LIB for each different population of noise transient. 
The different signal models for each glitch population can then be used for Bayesian model selection, which can determine the type of each new noise transient that is detected in the data.

For two competing models $M_{i}$ and $M_{j}$ the Bayes factor is given by the ratio of the evidences,

\begin{equation}
B_{i,j} = \frac{ p(D|M_{i}) }{ p(D|M_{j}) }, 
\label{Bayes}
\end{equation}
where $p(D|M_{i})$ is the evidence for model $M_{i}$ given data D, and $p(D|M_{j})$ is the evidence for model $M_{j}$ given data D \cite{Sivia1996}. The evidence for each model is calculated by integrating the likelihood $p(D|\theta ,M)$, multiplied by the prior $p(\theta |M)$, for all parameter values $\theta$. 

\begin{equation}
p(D|M) = \int_{\theta} p(\theta |M) p(D|\theta ,M) d\theta.
\label{EvidInt}
\end{equation}
The prior represents what we already know before any analysis of the data. The likelihood includes new information from the data.
First we calculate a signal vs. noise Bayes factor $B_{S,N}$. Taking the log of the Bayes factor gives

\begin{equation}
log(B_{S,N}) = log[p(D|M_{S})] - log[p(D|M_{N})] ,
\end{equation}
where $M_{S}$ and $M_{N}$ are our signal and noise models. To compare two different glitch models, $M_{type1}$ and $M_{type2}$, we can then subtract our signal vs. noise Bayes factors to obtain a new Bayes factor that determines the glitch type

\begin{equation}
log(B_{type1,type2}) = log(B_{S_{type1},N}) - log(B_{S_{type2},N}).
\end{equation}
For a large number of model parameters the evidence integral becomes difficult. This problem is solved using nested sampling, a description of which is given else where \cite{Sivia1996, 2010PhRvD..81f2003V}. The nested sampling algorithm produces posterior distributions for the values of the PC coefficients. 

A model for a type of noise transient is considered to be correct if log $B_{type1,type2} > 10$. We choose 10 to be conservative as a Bayes factor obtained by running on random noise can vary by around 5. A flat, uniform prior is used for the PC coefficients for each transient type. To calculate the minimum and maximum values for the PC coefficient priors we use the method described by Logue {\it et.\ al.} \cite{2012PhRvD..86d4023L} of projecting the transient waveforms on to the PCs. A Gaussian likelihood function is used, which is described in the LALInference paper by Veitch {\it et.\ al.} \cite{2014arXiv1409.7215V}. We choose the number of PCs that account for a large percentage ($\gtrsim70\%$) of the variance of the dataset.

For glitches in real detector noise a trigger generator will be used before running LIB. For this study we take the GPS times for the events from the log file that is produced when simulating the noise transients.

%%%%%%%%%%%%%%%%%%%%%%%%%%%%%%%%%%%%%%%%%%%%%%%%%%%%%%%%%%%%%%%%%%%%%%%%%%%%%%%%%%%%%%%%%%%%%%%%%%%%%%%%%%%

\subsection{Wavelet Detection Filter and Machine Learning}

WDF-ML consists of a event detection algorithm, Wavelet Detection Filter (WDF), followed by a Machine Learning (ML) classification procedure. 
WDF is part of the Noise Analysis Package (NAP), a C++ library embedded in python, developed by the Virgo Collaboration~\cite{NAP}.

\subsubsection{Wavelet Detection Filter.}

Wavelet-based algorithms are well tuned for the identification of noise transients because they
decompose the data into multiple time-frequency resolution maps. The
efficiency in detecting transients is linked to the similarities between the analyzing wavelet and the waveforms of the transients.  
As different wavelet types could better match different waveform morphologies, WDF performs wavelet domain decomposition using different types of wavelet basis, including the Daubechies and Haar wavelets \cite{daubechies1992ten, Mallat}.

A wavelet transform is similar to a Fourier transform. The Fourier transform sinusoidal waves are replaced
by an orthonormal basis generated by translations (shifting) and dilations (scaling) of the mother wavelet 

\begin{equation}
	\psi_{a,b}(t) = \frac{1}{\sqrt{b}}\psi\left(\frac{t-a}{b}\right)\label{eq:wavelet},
\end{equation}
where $b$ is the scale and $a$ is the translation. 
The wavelet transform of a signal $f(t)$ is defined as the projection of $f$ on the wavelet basis

\begin{equation}
Wf(a,b)=\left<f,\psi_{a,b}\right>=\int_{-\infty}^{+\infty}f(t)\frac{1}{\sqrt{b}}\psi^{*}\left(\frac{t-a}{b}\right)\ \ dt,
\label{eq:wavtransf}
\end{equation}
where $\psi^{*}$ is the complex conjugate of the mother wavelet. The wavelet transform has a time frequency resolution that depends
on the scale $b$. The time spread is proportional to $b$, and the
frequency spread is proportional to the inverse of $b$.
The discrete wavelet transform 
uses a discrete set of the wavelet scales and translations.
This transform decomposes the signal into a mutually orthogonal set
of wavelets.

\subsubsection{Data Conditioning and Trigger Detection.}
In Table \ref{codes} we outline the steps of the WDF-ML classification procedure.
The first five minutes of data are used to estimate the parameters for the following whitening filter in the time-domain. 
The whitening procedure is based on a Linear Predictor Filter, whose parameters are estimated through a parametric Auto Regressive (AR) model fit 
to the noise PSD, as described in \cite{cuoco-whitening}. One of the AR parameters is the standard deviation $\sigma$ of the background noise, which is used in the wavelet de-noising procedure.   

A signal $x_{i}$ that is corrupted by additive Gaussian
random noise $n_{i}\sim \mathcal{N}(0,\sigma^{2})$ is given by

\begin{equation}
%\[
x_{i}  =h_{i}+n_{i},\,\,\,\, i=0,1,...N-1 , 
%\]
\end{equation}
where $h_{i}$ is the transient signal.
The signal $x_{i}$ is used to find an approximation
$\hat{h}_{i}$ to the original $h_{i}$, which minimizes the mean squared
error

\begin{equation}
\Vert\mathbf{h}-\hat{\mathbf{h}}\Vert^{2}=\frac{1}{N}\sum_{i=0}^{N-1}|h_{i}-\hat{h}_{i}|^{2}.
\end{equation}
If an orthogonal wavelet transform $W$ is applied to the
sequence of data $x_{i}$, we obtain

\begin{equation}
W(x_{i})=W(h_{i})+W(n_{i}).
\end{equation}
For a given wavelet thresholding function $t$ the 
threshold based de-noising can be written as

\begin{equation}
\hat{h_{i}}=W^{-1}(t[W(x_{i})]).
\end{equation}
The thresholding function is applied to the wavelet transform of the noisy signal, then the output
is inverted and the wavelet transformed. The effectiveness of the technique is dependent upon
the choice of wavelet used, the decomposition level, and the amplitude of the threshold value. 

For a given threshold $t$ and wavelet coefficient $w$, the wavelet coefficient is retained if $|w|>t$,
or is set to zero if $|w|<t$.   
This removes wavelet coefficients that are due to background noise and retains wavelet coefficents that are due to the transient waveforms.
WDF uses the universal Donoho and Johnstone threshold method \cite{donoho1994ideal}, where

\begin{equation}
	t=\sqrt{2\log N}\hat{\sigma}\label{eq:dj},
\end{equation}
$N$ is the number of data points, and $\hat{\sigma}$ is an
estimate of the noise level $\sigma$, estimated during the AR parametric fit to the data. 

The wavelet coefficients contain the energy of the transient at different
scales. After the wavelet thresholding procedure is applied, only the highest
coefficients of the wavelet transform remain. These coefficients are expected to contain
only features of the transient waveforms. 
The energy of the transient is given by the sum of the square of the coefficients above the threshold value. The SNR is then given by the energy
divided by $\hat{\sigma}$.

WDF outputs a list of triggers, which include the maximum SNR and frequency, a GPS starting time for the transient, the transient  
duration, the name of the wavelet family which triggered the event, and the full list of the wavelet coefficients after the de-noising procedure.
The peak frequency of the transient is estimated as

\begin{eqnarray}
	f_{max}=\frac{f_s}{2.0\times window}\times b,
\end{eqnarray}
where $f_s$ is the sampling frequency, $window$ is the window used in the WDF process, and $b$ is the scale of the 
wavelet transform corresponding to the coefficient with the maximum value.
The event duration is estimated after applying a clustering step for events that are closer than $0.01$ s. 
 
For WDF-ML to correctly identify the glitch, the choice of window size and overlapping parameter between two consecutive sliding windows becomes important. 
For this data set we select a window size of $1024$ points. As there is no re-sampling filter in the data pre-processing, the data is sampled at $16384$ Hz,
 therefore, with $1024$ points the time window is $0.0625$ seconds. This ensures that the waveforms of duration $2$ ms will be inside the window. 
An overlap value of $0.05$ seconds was used in order to avoid problems caused by a transient waveform being in two consecutive windows. 

%%%%%%%%%%%%%%%%%%%%%%%%%%%%%%%%%%%%%%%%%%%%%%%%%%%%%%%%%%%%%%%%%%%%%%%%%%%%%%%

\subsubsection{Machine learning classification.}

ML classification procedures can be supervised or unsupervised \cite{abu2012learning}. A supervised ML algorithm trains on a sample of correctly labelled data. An unsupervised classification procedure has no labelled training set of data. WDF-ML uses an unsupervised classification procedure, as we have no previously labelled data set on which to train the algorithm. A supervised ML procedure will be implemented in a future study using information from auxiliary monitoring channels \cite{2013PhRvD..88f2003B, 2013CQGra..30o5010E}. The unsupervised procedure consists of a clustering algorithm to identify classes of events in the parameter space  created by the wavelet decomposition. WDF-ML applies the same ML classification algorithm, GMM, as described in section~\ref{sec:PCAT-classification}, but other clustering algorithms could be used, such as Affinity Propagation~\cite{frey2007clustering} or Kmeans~\cite{Arthur:2007qy}.
 
Wavelet coefficients are computed from the triggers and stored in an $n\times m$ matrix, 
where $n$ is the number of triggers, and $m$ is the length of the wavelet window. Most of the matrix elements are zeros,
as they are the coefficients that do not pass the thresholding step.
Dimensional reduction is required to retain the most important features of the matrix. This is achieved by first applying PCA, which reduces $m$ to $10$, and then projecting the remainder of the coefficients
on a two-dimensional space with Spectral Embedding \cite{Ng01onspectral, belkin2003laplacian}. 
Spectral Embedding finds a
low-dimensional representation of the data using a spectral decomposition of the graph Laplacian. 
The GMM ML algorithm is then applied to the reduced coefficients for classification.

%%%%%%%%%%%%%%%%%%%%%%%%%%%%%%%%%%%%%%%%%%%%%%%%%%%%%%%%%%%%%%%%%%%%%%%%%%%%%%%
%%%%%%%%%%%%%%%%%%%%%%%%%%%%%%%%%%%%%%%%%%%%%%%%%%%%%%%%%%%%%%%%%%%%%%%%%%%%%%%
%%%%%%%%%%%%%%%%%%%%%%%%%%%%%%%%%%%%%%%%%%%%%%%%%%%%%%%%%%%%%%%%%%%%%% 
%%%%%%%%%%%%%%%%%%%%%%%%%%%%%%%%%%%%%%%%%%%%%%%%%%%%%%%%%%%%%%%%%%%%%%%%%%%%%%%

\section{Data Sets}
\label{dataset}

For the sake of this investigation, we assume all advanced detectors to be affected by the same populations of glitches. Thus we use early aLIGO sensitivity curves for the Livingston detector to generate simulated Gaussian noise \cite{2014ApJ...795..105S}. We do not use real data for this study because we need to know all of the properties of the transients in the data set in order to accurately test the different methods.
We generate three different data sets containing different types of simulated noise transients, which are added to the Gaussian noise in five 
second intervals. The three data sets are designed to test if the different algorithms can classify transients by frequency, SNR and waveform morphology. We consider three different waveform morphologies: sine Gaussian (SG), Gaussian (G) and Ring-down (RD). 

\subsubsection{Sine Gaussian.}

The Sine Gaussian waveforms are defined by, 

\begin{equation}
	h_{\times}(t)= h_{0} \sin[2\pi f_{0} (t-t_{0})]e^{-(t-t_{0})^{2}/2\tau^{2}},
	\label{SGeqn1}
\end{equation}

\begin{equation}
	h_{+}(t)= h_{0} \cos[2\pi f_{0} (t-t_{0})]e^{-(t-t_{0})^{2}/2\tau^{2}} ,
	\label{SGeqn2}
\end{equation}
where $\tau=Q/\sqrt{2}\pi f_0$, $f_0$ is the central frequency, $Q$ is the 
quality factor, $t_{0}$ is the GPS time at the centre of the sine Gaussian, 
and $h_{0}=\hbox{hrss}/\sqrt{\tau}$, where $\hbox{hrss}$ is the root sum 
squared amplitude of the transient. The $\tau$ parameter determines the width of the simulated waveform in the time-domain.

\subsubsection{Gaussian.}

The Gaussian simulated waveforms are defined by, 

\begin{equation}
	h_{\times}(t)= h_{0} e^{-(t-t_{0})^{2}/2\tau^{2}}, 
\label{Geqn1}
\end{equation}

\begin{equation}
	h_{+}(t)= h_{0} e^{-(t-t_{0})^{2}/2\tau^{2}}. 
\label{Geqn2}
\end{equation}
The Gaussian waveforms are centred at zero frequency with the maximum frequency determined by the duration. The spike glitches that were observed in S6 were characterized by a Gaussian waveform morphology in the time-domain \cite{2014arXiv1410.7764T}. Their characteristic time series contained a dip followed by an upwards spike that typically lasted for a few milliseconds.   

\subsubsection{Ring-down.}

The Ring-down simulated waveforms are defined by, 

\begin{equation}
	h_{\times}(t)= h_{0} \sin[2\pi f_{0} (t-t_{0})]e^{-(t-t_{0})/2\tau} ,
\label{RDeqn1}
\end{equation}
\begin{equation}
	h_{+}(t)= h_{0} \cos[2\pi f_{0} (t-t_{0})]e^{-(t-t_{0})/2\tau}, 
\label{RDeqn2}
\end{equation}
where $t>t_0$.
The Ring-down waveforms are similar to high SNR spike glitches, which were observed with time-domain waveforms that Ring-down after their initial spike \cite{2014arXiv1410.7764T}.

\subsection{Data Set 1}

The first data set contains 1000 simulated Gaussian  transients and 1000 simulated sine Gaussian transients of different duration, frequency and SNR. The  transient waveforms were generated with $Q$, hrss, duration and frequency values distributed uniformly between the minimum and maximum values, shown in Table A1.

\subsection{Data Set 2}

Data set 2 includes 1000 simulated sine Gaussian transients and 1000 Ring-down transients with SNR uniformly distributed between 1 and 400. All transients were generated with identical frequency (400 Hz) and duration (2 ms). This data set is designed to test that the different algorithms can classify transients by waveform morphology only.  

\subsection{Data Set 3}

Data set 3 includes 1000 Gaussian, 1000 sine Gaussian, and 1000 Ring-down transients. The waveform parameters in this data set have a large range of values, which makes this data set more challenging to classify than the first two data sets. The parameters of the simulated noise transients in this data set allow us to test the limitations of the three different classifying methods. The parameters for the simulated waveforms are distributed uniformly between the minimum and maximum values in Table A2. %\ref{Data1}.

\begin{figure}[!t]
\begin{centering}
	\includegraphics[width=7.5cm]{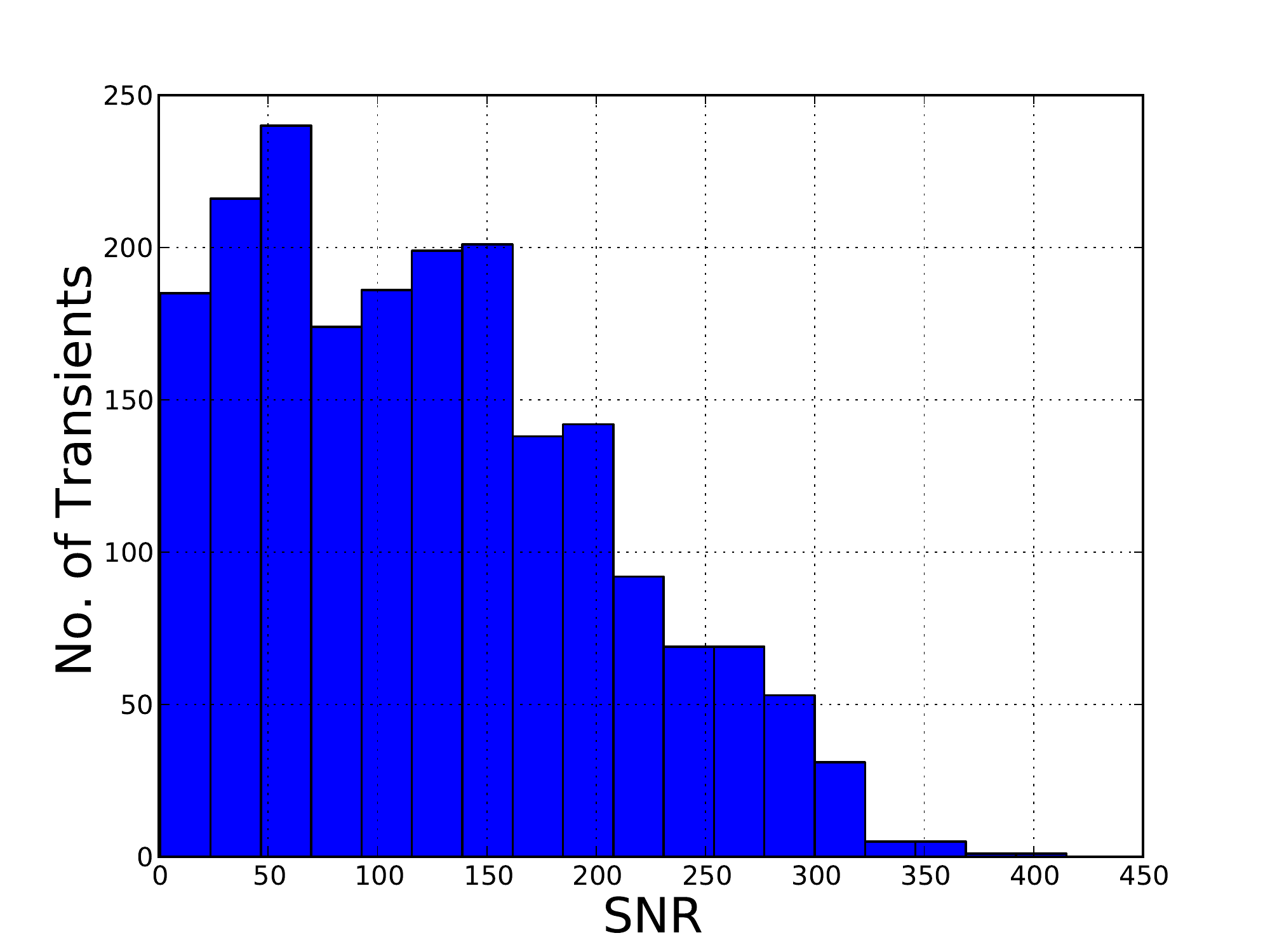}
        \includegraphics[width=7.5cm]{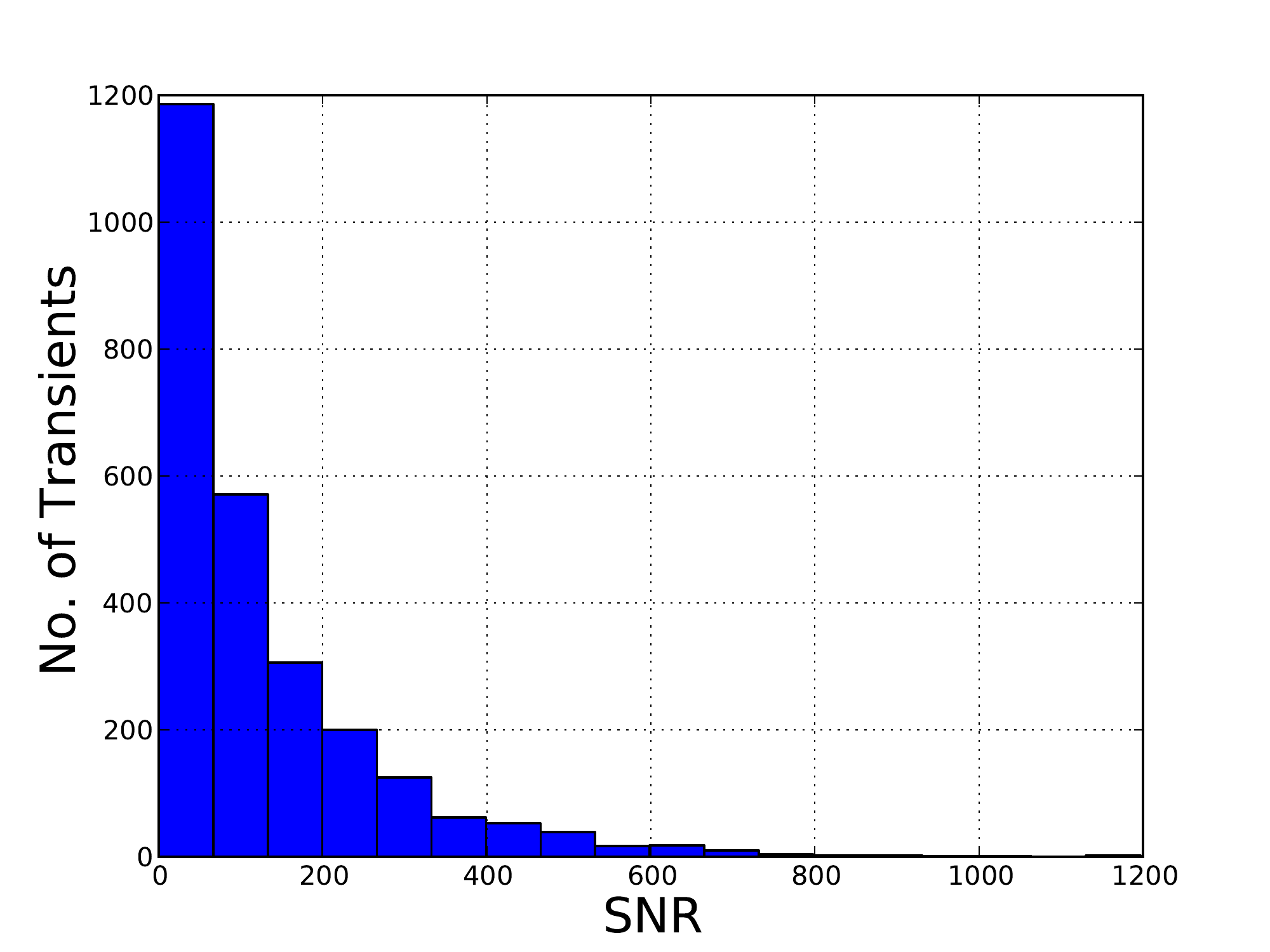}
	\caption{The left plot shows the SNR distribution for the data set 1 transients. 
The right plot shows the SNR distribution for the data set 3 transients.}
\label{fig:WDF-scatter-plots}
\end{centering}
\end{figure}

%%%%%%%%%%%%%%%%%%%%%%%%%%%%%%%%%%%%%%%%%%%%%%%%%%%%%%%%%%%%%%%%%%%%%%%%%%%%%%%
%%%%%%%%%%%%%%%%%%%%%%%%%%%%%%%%%%%%%%%%%%%%%%%%%%%%%%%%%%%%%%%%%%%%%%%%%%%%%%%
%%%%%%%%%%%%%%%%%%%%%%%%%%%%%%%%%%%%%%%%%%%%%%%%%%%%%%%%%%%%%%%%%%%%%%%%%%%%%%%
%%%%%%%%%%%%%%%%%%%%%%%%%%%%%%%%%%%%%%%%%%%%%%%%%%%%%%%%%%%%%%%%%%%%%%%%%%%%%%%%

\section{Results}
\label{Results}

\subsection{Data Set 1 Results}

In this subsection we show the results for classifying the two transient types in the first data set. 

\subsubsection{PCAT.}

PCAT identifies 1309/2000 transients above the threshold. The first five PCs account for $75\%$ of the variance in the data (Figure 1). The first twelve PCs describe all the major 
features of the glitches. 
Clustering with the first five PC coefficients leads to seven transient types, as shown in Table \ref{table:MDC3}. Types 2, 3, 5 and 6 
identify Gaussian transients. Types 1 and 7 identify sine Gaussian 
transients.

The breakdown of Gaussian and sine Gaussians in to multiple types can be 
understood as a separation in frequency and SNR, for the Gaussian and 
the sine Gaussian waveforms, respectively, as shown in Figure 5(a).
Types 3 and 6 are the lower frequency Gaussian waveforms $\sim (40-90)Hz$, and types 5 and 2 are the higher frequency Gaussian waveforms $\sim (100-150)Hz$. 
Type 7 contains, on average, sine Gaussian transients with SNRs larger by a factor of $\sim 5$, and a standard 
deviation larger by a factor of $\sim 10$, than type 1 transients. 

By forcing 
PCAT to cluster the data on a maximum of two types, 99\% of sine 
Gaussian and 100\% of Gaussian glitches are classified as type 1 and type 
2, respectively. 
The few misclassified glitches in this case correspond to transients whose identified GPS time is not correctly aligned with the peak of transient. This issue can be resolved by further tuning of the PCAT trigger generator. 

\begin{table}
\begin{centering}
\begin{tabular}{| c | c | c | c |}\hline
               &   SG      &   G      \\\hline

PCAT Type 1    &   8.5   &   0    \\
PCAT Type 2    &   0     &   15.4 \\
PCAT Type 3    &   0     &   19.5 \\
PCAT Type 4    &   0.9   &   0.2  \\
PCAT Type 5    &   0     &   35.9 \\
PCAT Type 6    &   0     &   29.0 \\
PCAT Type 7    &   90.5  &   0    \\\hline\hline

(maxcluster 2) PCAT Type 1    &     99  &  0 \\
(maxcluster 2) PCAT Type 2    &     1   &  100 \\\hline\hline

LIB  Type 1    &   99.9  &   5    \\
LIB  Type 2    &   0.1   &   95   \\\hline\hline

WDF Type 0    &   99.5  &   2.4     \\
WDF Type 1    &   0.3   &   46.1  \\
WDF Type 2    &   0.2  &   51.5   \\

    \hline
\end{tabular}
\caption{The table shows the LIB, PCAT and WDF-ML results for data set 1. The values show the percentage of the different morphologies classified in each type. The total number of simulated waveforms was 1000 of each type. The total number of glitches analysed were 1309 for PCAT, 1452 for LIB and 1814 for WDF-ML. }
\label{table:MDC3}
\end{centering}
\end{table}

\subsubsection{LIB.}

1452/2000 transients were large enough to be detected by LIB. Most of the detected transients had an SNR larger than 10. 7 PCs were used to classify the glitches in to two different types. The 7 type 1 PCs represented 97\% of the variance of the type 1 transients, and the 7 type 2 PCs represented 70\% of the variance of the type 2 transients. Although setting a threshold on $v$ suggests that seven is an ideal number of PCs, plotting the PCs shows that after the 5th PC, the rest consist of noise only, and do not contain any more information about the waveforms.  

The results are shown in Table \ref{table:MDC3}. LIB classified all of the glitches with a very high efficiency $(\geq 95\%)$. Type 1 is the main type for the sine Gaussian waveforms, and type 2 is mainly Gaussian waveforms. The 5\% of Gaussian waveforms that were in the incorrect class had low SNR values $(\leq 20)$. 

The Bayes factors that were obtained for all of the detected waveforms in this data set are shown in Figure \ref{fig:LIB-MDC3-Bayes}. If the type 1 waveforms have been correctly classified then the glitch type Bayes factor should be positive, and if the type 2 waveforms have been correctly classified then then the glitch type Bayes factor should be negative. When using the correct transient waveform model the increase in the log signal to noise Bayes factors is proportional to the square of the SNR. When using the incorrect transient waveform model the log signal to noise Bayes factors remain low as the SNR values of the transients increase. There is a clear difference in the log Bayes factors between the two types once the SNR becomes larger than 20.

\begin{figure}[!t]
\begin{centering}
	\includegraphics[width=7.5cm]{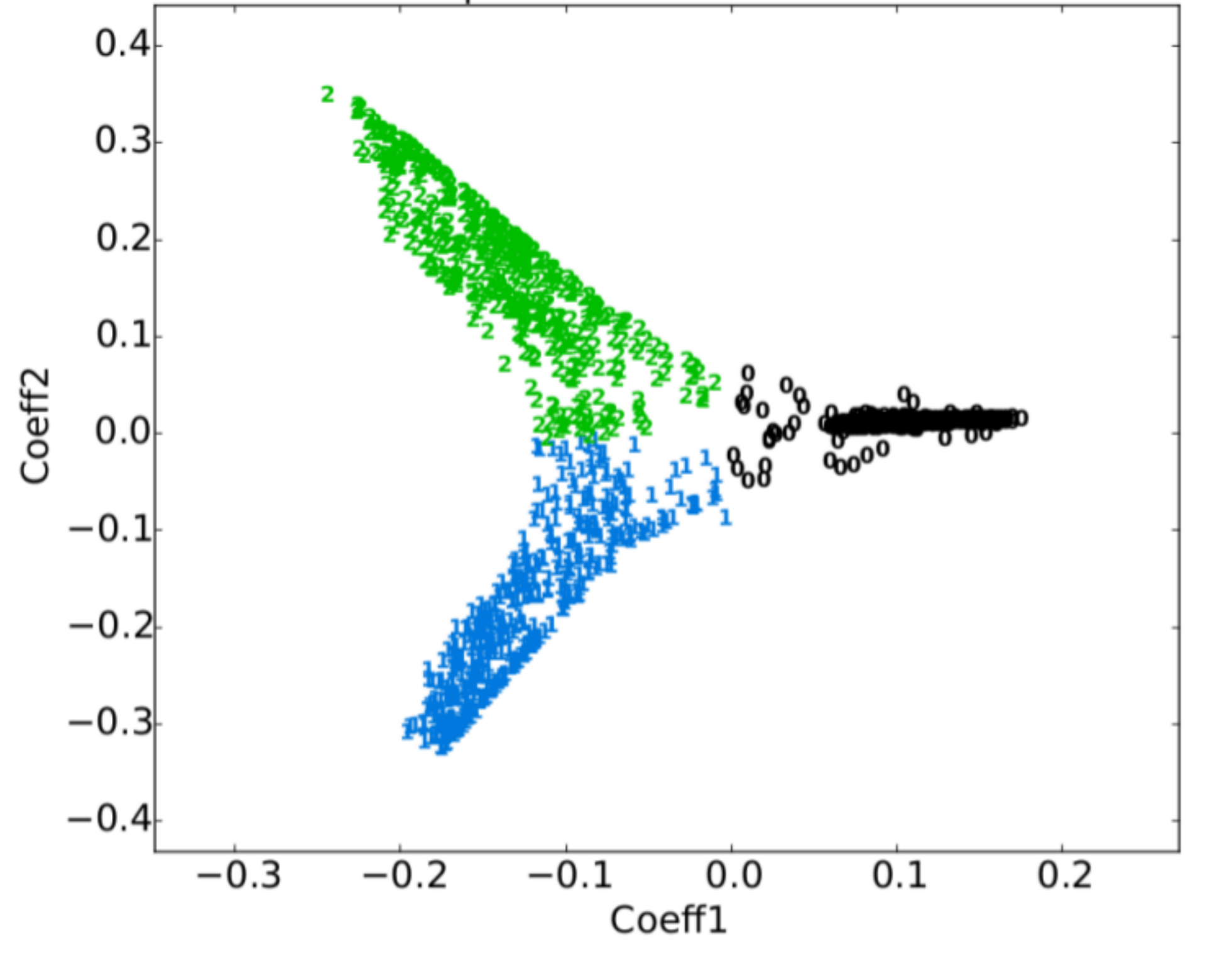}
        \includegraphics[width=7.5cm]{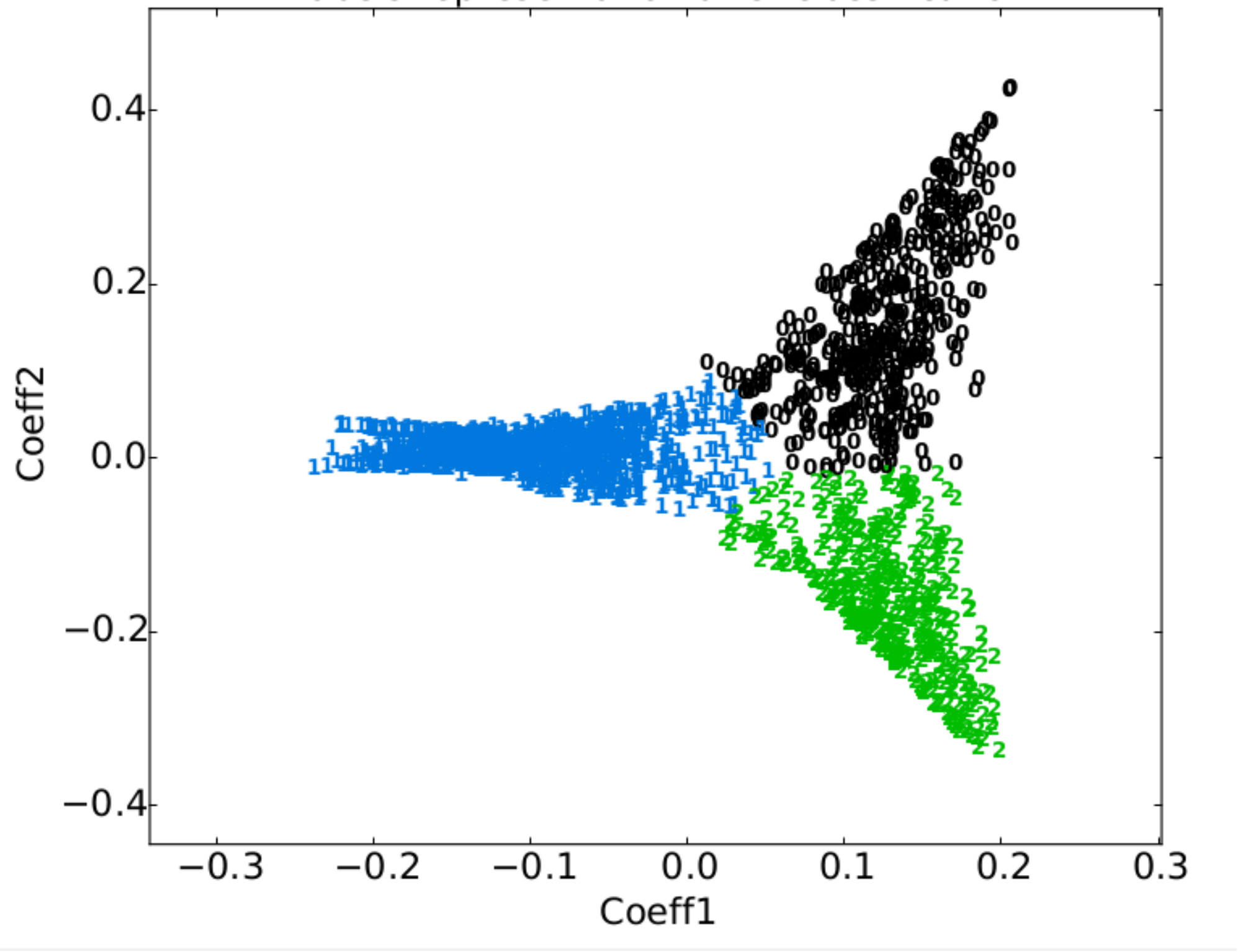}
	\caption{The left plot shows the values for the transformed and reduced wavelet coefficients obtained by WDF on the data set 1 transients. 
The right plot shows the transformed and reduced wavelet coefficients for the data set 2 transients. The two coefficients for the different types of transients are well separated in the parameter space. Wavelet coefficients are different for different data sets.}
\label{fig:WDF-scatter-plots}
\end{centering}
\end{figure}

\subsubsection{WDF-ML.}

WDF detected 1814/2000 transients using an SNR threshold of 15. 
The 10 PCs used represented $\sim95\%$ of the variance of the wavelet coefficients.
The results are shown in Table \ref{table:MDC3}. The efficiency for correct classification was 
higher than 97\% for both data sets.
Figure \ref{fig:WDF-scatter-plots} shows the ML results for the three types of transients found in the data.
The wavelet coefficients for different types of transients are well separated in the parameter space.
Type 0 contains the sine Gaussian waveforms. The Gaussian waveforms have been split in to two sub-types 1 and 2, where
The type 2 contains more lower SNR Gaussian waveforms, SNR between 25 and 150, than type 1. 

\subsubsection{Summary.}

We have found that all three methods have a very high efficiency for the correct classification of different types of noise 
transients. 
LIB and PCAT require that the number of glitch types are specified in advance. If the number of glitch types requested by PCAT is higher than the  
actual number of glitch types in the data set, then the waveforms will be classified by waveform morphology first, and then split in to further sub-types by 
frequency and SNR. WDF-ML has also shown that if it identifies more types than those present in the data, then the waveform morphologies will be split into further sub-types by SNR or frequency.  

As LIB needs a set of PCs in advance to create a signal model, it is only possible for LIB to classify known types of transients in the data. 
A new set of PCs will need to be created any time that a new family of glitches appears in the data. 
As PCAT and WDF-ML do not need any information about a glitch type before they start the classification procedure, 
they can begin to classify new transient types as soon as they appear in the advanced detector data. 
As LIB runs on one second of data at a time, when analyzing real glitches there may be multiple glitches of different types inside that one second of data, which could affect the efficiency of the classification. Multiple glitches in a small segment of data could also create problems during the windowing stage of WDF.

%%%%%%%%%%%%%%%%%%%%%%%%%%%%%%%%%%%%%%%%%%%%%%%%%%%%%%%%%%%%%%%%%%%%%%%%%%%%%%%
%%%%%%%%%%%%%%%%%%%%%%%%%%%%%%%%%%%%%%%%%%%%%%%%%%%%%%%%%%%%%%%%%%%%%%%%%%%%%%%

\subsection{Data Set 2 Results}

This subsection describes the results for the second data set, which is designed to test if the classification methods can classify noise transients by waveform morphology only. 

\begin{table}
\begin{centering}
\begin{tabular}{| c | c | c | c |}\hline
                    &   SG       &   RD    \\\hline

(94 PCs) PCAT Type 1     &    0.16    &   0  \\
(94 PCs) PCAT Type 2     &    18.0  &   18.6  \\
(94 PCs) PCAT Type 3     &    37.0  &   28.4  \\
(94 PCs) PCAT Type 4     &    0.16    &   0  \\
(94 PCs) PCAT Type 5     &    16.32  &   18.1  \\
(94 PCs) PCAT Type 6     &    0.16    &   0  \\
(94 PCs) PCAT Type 7     &    28.2  &   34.8  \\\hline\hline

(5 PCs) PCAT Type 1 & 0.32 & 12.6  \\
(5 PCs) PCAT Type 2 & 25.5 & 0.2   \\
(5 PCs) PCAT Type 3 & 20.4 & 1.3   \\
(5 PCs) PCAT Type 4 & 1.3 & 2.8   \\
(5 PCs) PCAT Type 5 & 0 & 37.4     \\
(5 PCs) PCAT Type 6 & 0 & 30.0     \\
(5 PCs) PCAT Type 7 & 52.4 & 0    \\
(5 PCs) PCAT Type 8 & 0 & 16.1    \\\hline\hline

(5 PCs, maxcluster 2) PCAT Type 1 & 1.1 & 97.4  \\
(5 PCs, maxcluster 2) PCAT Type 2 & 98.9 & 2.5   \\\hline\hline

LIB Type 1          &   97.8   &  4.8  \\
LIB Type 2          &   2.2    &  95.2 \\\hline\hline

WDF-ML Type 0       &   8.7  &  100  \\
WDF-ML Type 1       &   48.0  &  0  \\
WDF-ML Type 2       &   43.3  &  0  \\

    \hline
\end{tabular}
\caption{The table shows the results for LIB, PCAT and WDF-ML, for data set 2, which is designed to see if the methods can classify glitches by waveform morphology only. The values show the percentage of the different morphologies classified in each type. Two sets of PCAT results are included with different numbers of max clusters. The total number of glitches analysed were 1265 for PCAT, 1925 for LIB and 1914 for WDF-ML. }
\label{table:MDC4}
\end{centering}
\end{table}

\subsubsection{PCAT.}  
PCAT identifies 1265/2000 transients above the threshold. $94$ PCs account for $75\%$ of the variance of the waveforms. Clustering 
using the first $94$ PC coefficients results in seven different transient types, shown in Table \ref{table:MDC4}, of which three types only contain one low SNR transient.
Morphology classification is mixed: most types contain a roughly equal number of sine Gaussian and Ring-down transients. Transients are classified according to SNR, as shown in Figure \ref{fig:PCAT-MDC4-explained-variance}. This is due to after the 10th PC, the PCs only account for noise, and including too much noise degrades the efficiency of the classification algorithm. 

Changing the number of PCs used to $5$, the location of the ``knee'' of the variance curve (accounting for $51\%$ of the variance), yields better 
classification efficiency. Transients are first classified by waveform morphology and then broken down in to subclasses with different SNRs. The sine Gaussian waveforms are in types 2, 3 and 7. The Ring-down waveforms are contained in types 1, 5, 6 and 8. Type 4 contains less than 30 waveforms that are a mixture of the two types. The results show that PCAT is able to classify transients, by waveform morphology alone, with a very high efficiency when noisy PCs are not included.

The results can be improved further by limiting the maximum number of clusters to two. Type 1 contains the Ring-down waveforms, and type 2 contains the sine Gaussian waveforms. In this case, the few mis-classified transients either have low SNR ($\sim\!\!10$) or have waveforms with peaks that are not aligned with the GPS time for the transient.

\subsubsection{LIB.}

\begin{figure}[!t]
\begin{centering}
	\includegraphics[width=7.5cm]{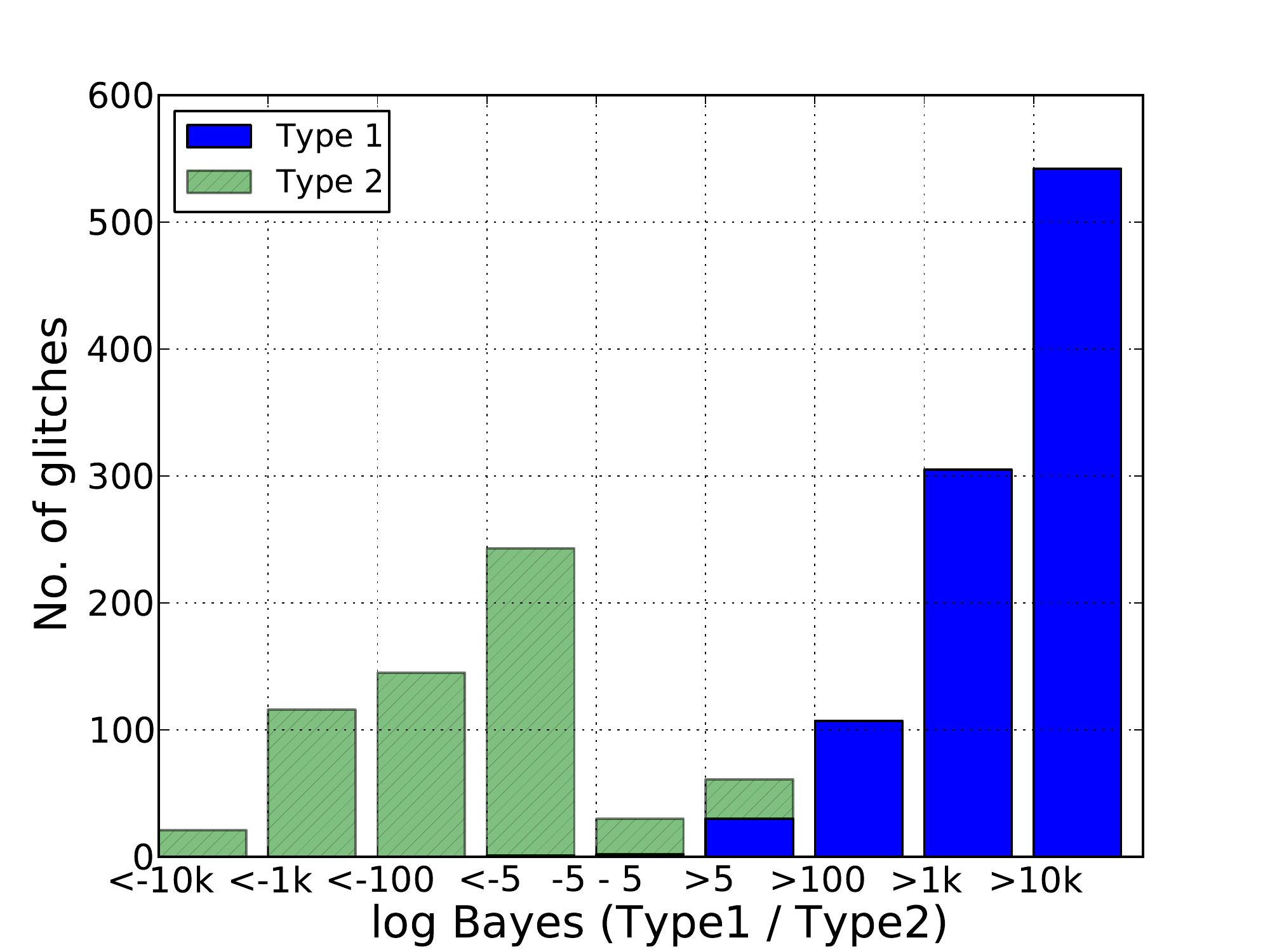}
        \includegraphics[width=7.5cm]{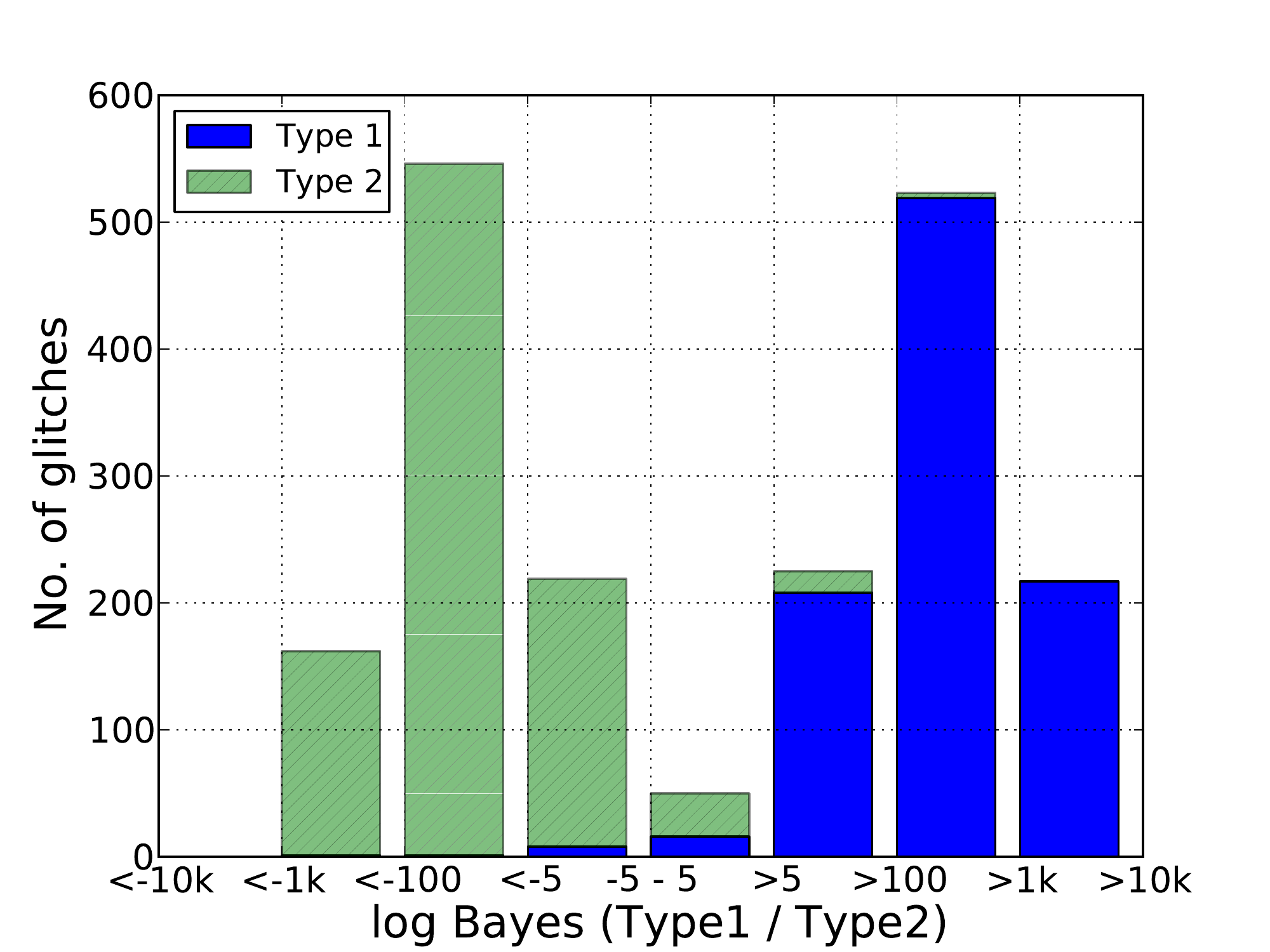}
	\caption{The histogram of the log Bayes factors are shown for all the the detected transients. The left plot shows the Bayes factors for data set 1, and right plot shows the Bayes factors for data set 2. Bayes factors are larger for noise transients with a higher SNR. }
\label{fig:LIB-MDC3-Bayes}
\end{centering}
\end{figure}

1925/2000 of the simulated waveforms were classified by LIB, as shown in Table \ref{table:MDC4}. 97.8\% 
of the transients with a sine Gaussian morphology were identified as type 1, and 95.2\% of the transients with a Ring-down morphology were 
classified as type 2 transients. LIB was clearly able to classify the transients by waveform morphology alone with a high efficiency. 
The simulated transients that were incorrectly classified by LIB had an SNR less than 20. 
The log Bayes factors for the two types of transients are shown in Figure \ref{fig:LIB-MDC3-Bayes}. The similar size and shape in distribution of Bayes factors shows that both of the transient types have the same distribution of SNR values. 

$7$ PCs were used for each transient type, as 7 PCs represented 90\% of the variance of the type 1 transients, and 80\% 
of the variance of the type 2 transients. As we know that the waveforms in each type are identical in this data set only one PC should be necessary to represent all of the variance of the waveforms. Plotting the variance curve showed a larger number of PCs were needed to accurately represent the data set. This is because the variance curve is affected by the noise included in the waveforms used to make the curve.
The PCs may give a better representation of the features of the transients if only high SNR transients are selected when creating the PCs. However, this may not always be possible if the transients do not occur at high SNR values. 

\subsubsection{WDF-ML.}

The 10 PCs represented $\sim\!98\%$ of the variance of the wavelet coefficients. 
WDF detected 1914/2000 transient waveforms with an SNR threshold $>15$, as shown in Table~\ref{table:MDC4}. WDF-ML was able to 
classify different noise transients by waveform morphology alone, with a high efficiency. The results for the ML procedure applied to the reduced coefficients is shown in Figure \ref{fig:WDF-scatter-plots}. There is a clear separation in the parameter space for the three different types. All of the detected Ring-down transients are in type 0. 
The sine Gaussian transients have been split in to two classes, which are type 1 and 2. The two types of sine Gaussian waveforms were not split by frequency or SNR in this case. The sine Gaussian and Ring-down waveforms can be easily incorrectly classified with a wrong choice of overlap value and window size, because if the waveform is split over two consecutive analyzing windows then a sine Gaussian would be cut off in the middle of the waveform, which would make it appear to be a  Ring-down waveform. In real data most glitches have a duration of a few milliseconds, therefore a window of a few 100 milliseconds will be used.
 
\begin{figure}[!t]      
        \subfigure[]{\label{fig:a}\includegraphics[width=7.8cm]{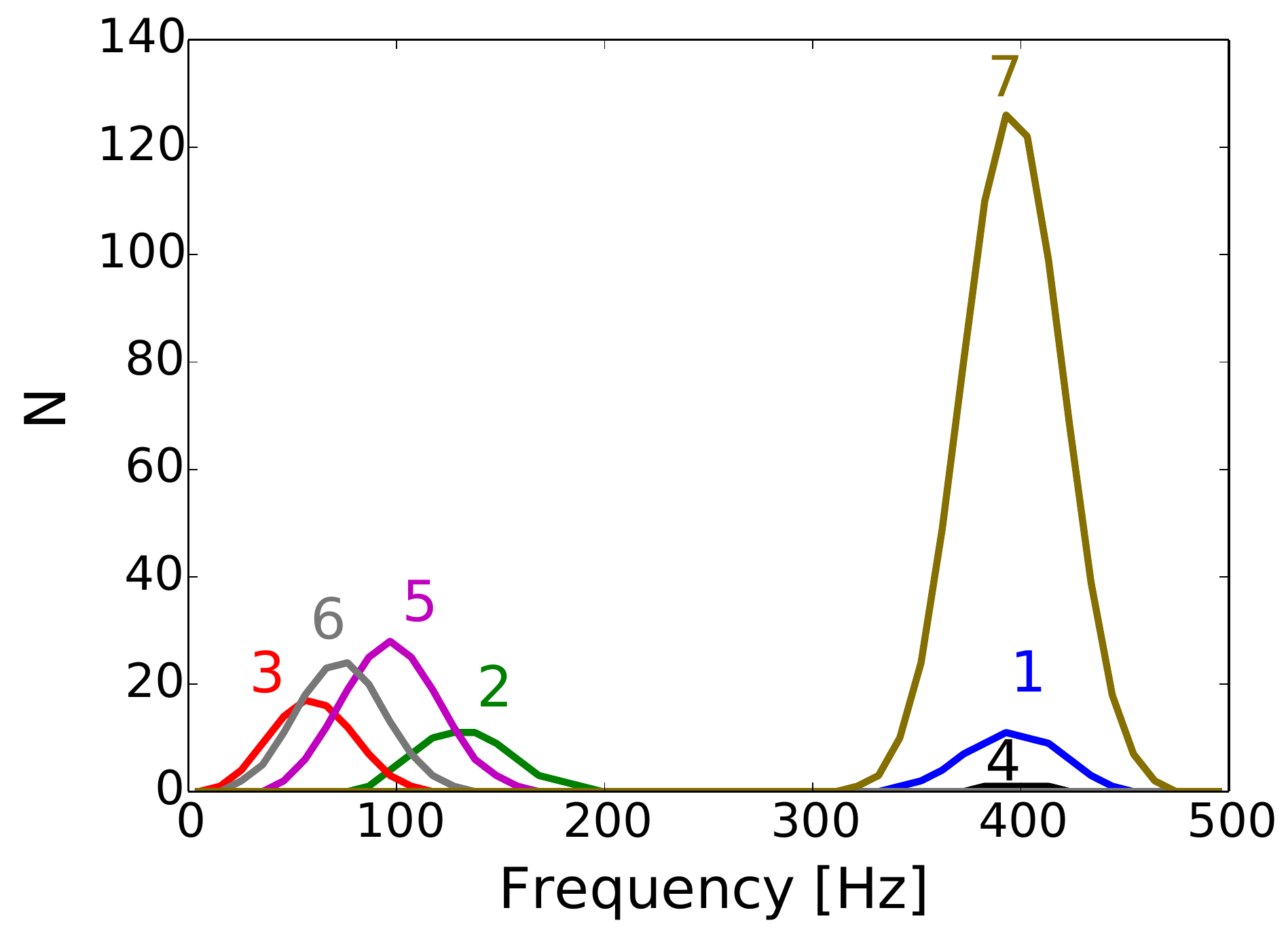}}       
        \subfigure[]{\label{fig:b}\includegraphics[width=7.8cm]{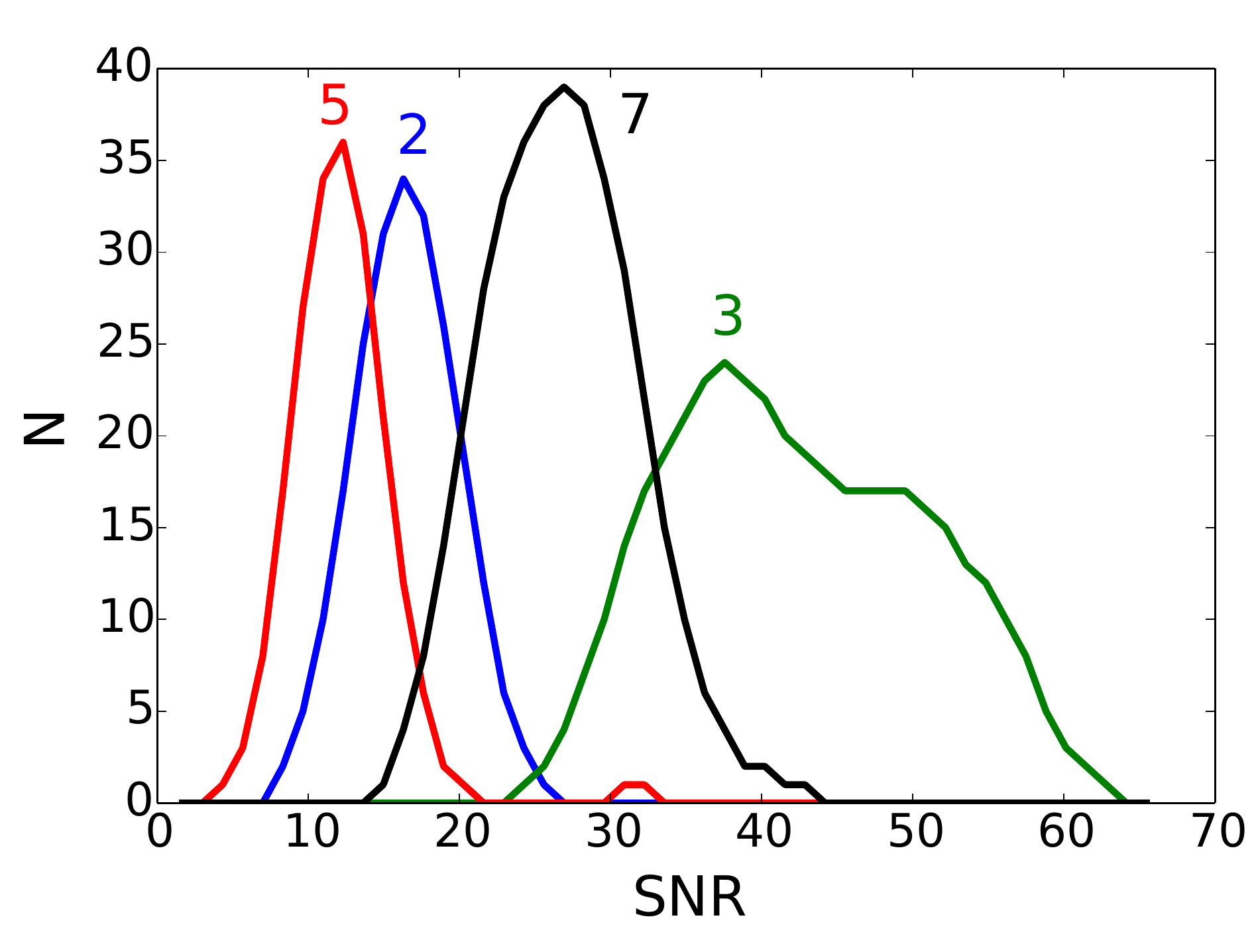}}
	\caption{(a) shows the PCAT reconstructed frequency distributions of the different transient types in data set 1. Transients are classified by waveform morphology first and then split into further sub types by frequency or SNR. (b) shows the PCAT reconstructed SNR distributions for the transients in data set 2 when using $94$ Principal Components. Numbers identify glitch types.} 
\label{fig:PCAT-MDC4-explained-variance}
\end{figure}   

\subsubsection{Summary.}

We have shown that all three classification methods are able to classify transient waveforms by morphology alone, with a very high efficiency. 
Depending on the maximum number of allowed classes, PCAT may classify transients not only by morphology, but also by SNR, assigning SNR sub-classes to each transient morphology. Including too many PCs degrades classification efficiency, because too much noise is included. For PCAT the most effective method for the selection of the number of PCs was found to be the number suggested by the position of the knee of the variance curve. Choosing a high percentage of the variance may not be ideal in the case of glitches because it is not possible to eliminate background noise from the glitch waveforms. 

%%%%%%%%%%%%%%%%%%%%%%%%%%%%% data set 3 table
\begin{table}[!t]
\begin{centering}
\begin{tabular}{| c | c | c | c |}\hline
                     &  SG       &  G      &   RD      \\\hline
PCAT (33PCs) Type 1  &   16.9  &   0   &   14.5  \\
PCAT (33PCs) Type 2  &   4.8   &   100 &   9.6   \\
PCAT (33PCs) Type 3  &   37.1  &   0   &   41.8  \\
PCAT (33PCs) Type 4  &   10.7  &   0   &   4.5   \\
PCAT (33PCs) Type 5  &   4.5   &   0   &   0.7   \\
PCAT (33PCs) Type 6  &   21.2  &   0   &   19.7  \\
PCAT (33PCs) Type 7  &   4.8   &   0   &   9.2   \\\hline\hline

PCAT (20PCs) Type 1  &   15.5  &   0   &   13.6  \\
PCAT (20PCs) Type 2  &   36.8  &   0   &   41.4  \\
PCAT (20PCs) Type 3  &   14.2  &   0   &   13.0  \\
PCAT (20PCs) Type 4  &   9.1   &   0   &   13.0  \\
PCAT (20PCs) Type 5  &   0.8   &   0   &   0.3   \\
PCAT (20PCs) Type 6  &   21.8  &   0   &   17.2  \\
PCAT (20PCs) Type 7  &   1.8   &  100  &   1.5   \\\hline\hline

LIB (5PCs) Type 1    &   39.5  &  4.9  &   23.8  \\
LIB (5PCs) Type 2    &   17.3  &  88.3 &   23.2  \\
LIB (5PCs) Type 3    &   43.3  &  6.8  &   53.0  \\\hline\hline

WDF-ML Type 1        &   89.5  &  9.6 &  86.9  \\
WDF-ML Type 2        &   5.9   &  49.7 &  7.0     \\
WDF-ML Type 3        &   4.6   &  40.7 &  6.1   \\
 
\hline
\end{tabular}
\caption{The table shows the PCAT, LIB and WDF-ML results for data set 3. The values show the percentage of the different morphologies classified in to each type. Two sets of PCAT results are included with different numbers of maximum PCs. The total number of glitches analysed were 1480 for PCAT, 2162 for LIB and 2547 for WDF-ML. All methods were unable to distinguish between the sine Gaussian and Ring-down waveform morphologies in this data set.}
\label{table:MDC1}
\end{centering}
\end{table} 

%%%%%%%%%%%%%%%%%%%%%%%%%%%%%%%%%%%%%%%%%%%%%%%%%%%%%%%%%%%%%%%%%%%%%%%%%%%%%%%%%%%%%%%%%%%
%%%%%%%%%%%%%%%%%%%%%%%%%%%%%%%%%%%%%%%%%%%%%%%%%%%%%%%%%%%%%%%%%%%%%%%%%%%%%%%%%%%%%%%%%%%%

\subsection{Data Set 3 Results}
This subsection shows the results for the third data set, which contains transient waveforms that have a wide range of parameters. 

\subsubsection{PCAT.}

PCAT identifies 1480/3000 of the noise transients. They are classified into seven different types, as shown in Table \ref{table:MDC1}. 33 PCs represented 75\% of the variance of the data set. The classification results are mixed, with type 2 being the exception, containing $100\%$ of the simulated transients belonging to the Gaussian morphology. From the distribution of transient peak frequencies for each PCAT type, shown in Figure 6(a), the mixed-classification can be understood as a frequency-based classification. Type 3 contains the highest frequency transients. Type 7 and 5 contain the lower frequency transients. 
There are a few Ring-down and sine Gaussian transients that are classified as type 2 (G), which have frequency distributions similar to the Gaussian waveforms ($70-150$ Hz). 
The wide range of parameters of the simulated waveforms makes it hard to capture the full range of the waveform parameters in the first few PCs, therefore, the main parameter captured by the PCs is frequency, on which the classification is then based.

Table \ref{table:MDC1} also shows the results using 20 PCs, which corresponds to the approximate location of the knee of the variance curve. 
Changing the method used to select the number of PCs that represent this data set did not lead to an improvement in the result.

%%%%%%%%%%%%%%%%%%%%%%%%%%%%%%%%%%%%%%%%%%%%%%%%%%%%%%%%%%

\subsubsection{LIB.}

Using 5 PCs 2162/3000 of the transient waveforms were classified by LIB. The 5 PCs represent 67\% of the variance of the sine Gaussian waveforms, 93\% of the variance of the Gaussian waveforms and 80\% of the variance of the Ring-down waveforms. The results are shown in Table \ref{table:MDC1}. The table shows that type 2 contains the majority of the Gaussian waveforms, and the other two types of waveform morphologies are mixed in types 1 and 3. Figure 6(b) shows the frequency distribution of the three different types of transients. Type 1 contains the mid frequency range (300-700 Hz) waveforms and type 3 contains higher frequency waveforms (700-1500 Hz). A small number of low frequency sine Gaussian and Ring-down morphologies $(\sim 20\%)$ were in the type 2 class with the Gaussian transients. The 12\% of Gaussian morphologies that were incorrectly classified had low SNR values $(\lesssim 20)$. 

The frequency distribution for the total number of simulated transients in this data set is uniform. However, as only a small number of the total waveforms were used in making the PCs used to create the signal model, the frequency distributions for the transient waveforms used to make the PCs was not uniform. The type 1 (SG) transients used to make the PCs contained more mid frequency range waveforms. The type 3 (RD) transients used to make the PCs contained more higher frequency waveforms. This shows that for real glitch types with a wider range of parameters, we need to be careful in the selection of waveforms that are used to make the PCs for the signal model, so that we do not introduce a bias in the results in certain areas of the parameter space. 

\begin{figure}[!t]
	\subfigure[]{\label{fig:a}\includegraphics[width=5.30cm]{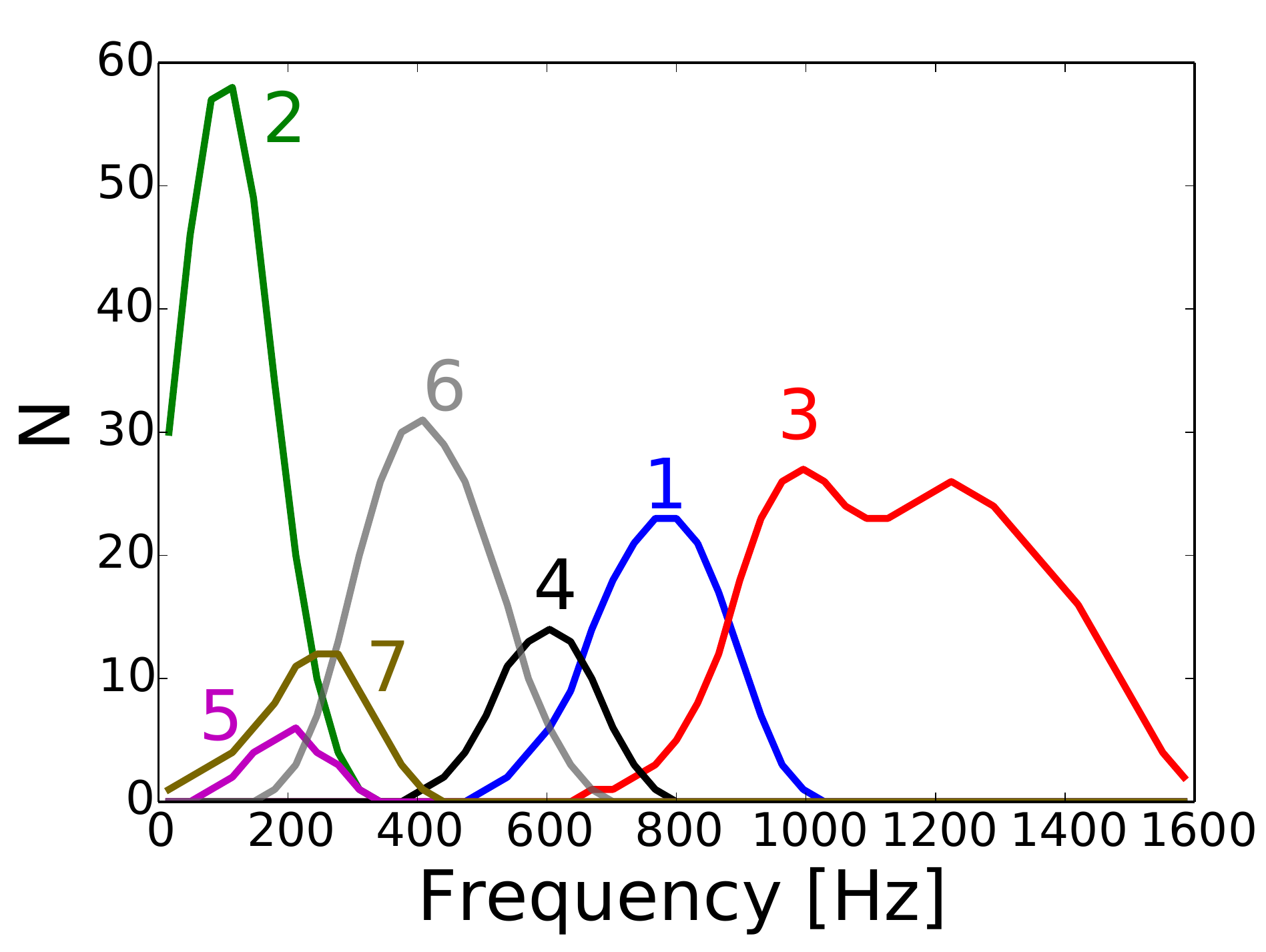}}
        \subfigure[]{\label{fig:b}\includegraphics[width=5.30cm]{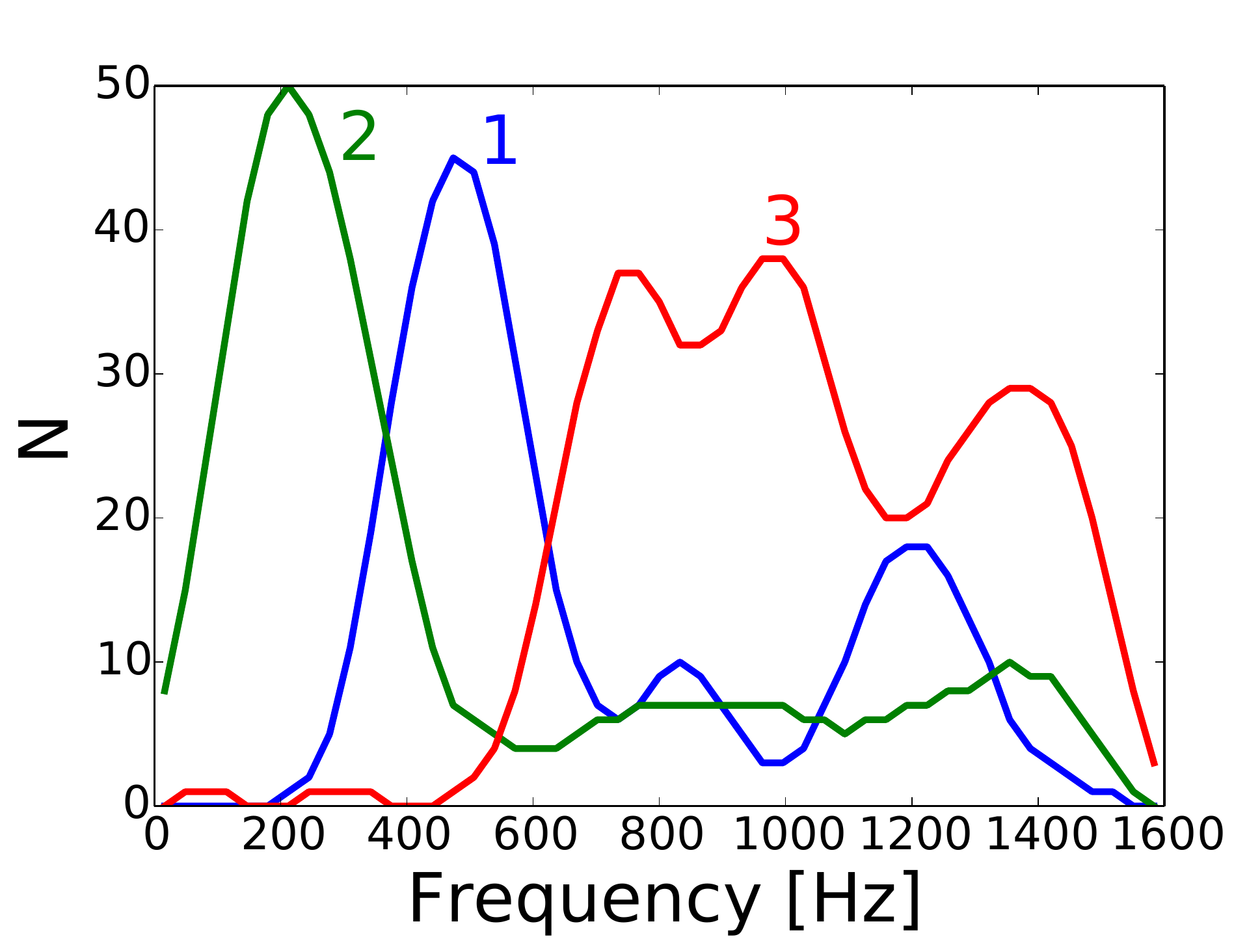}}
        \subfigure[]{\label{fig:c}\includegraphics[width=5.30cm]{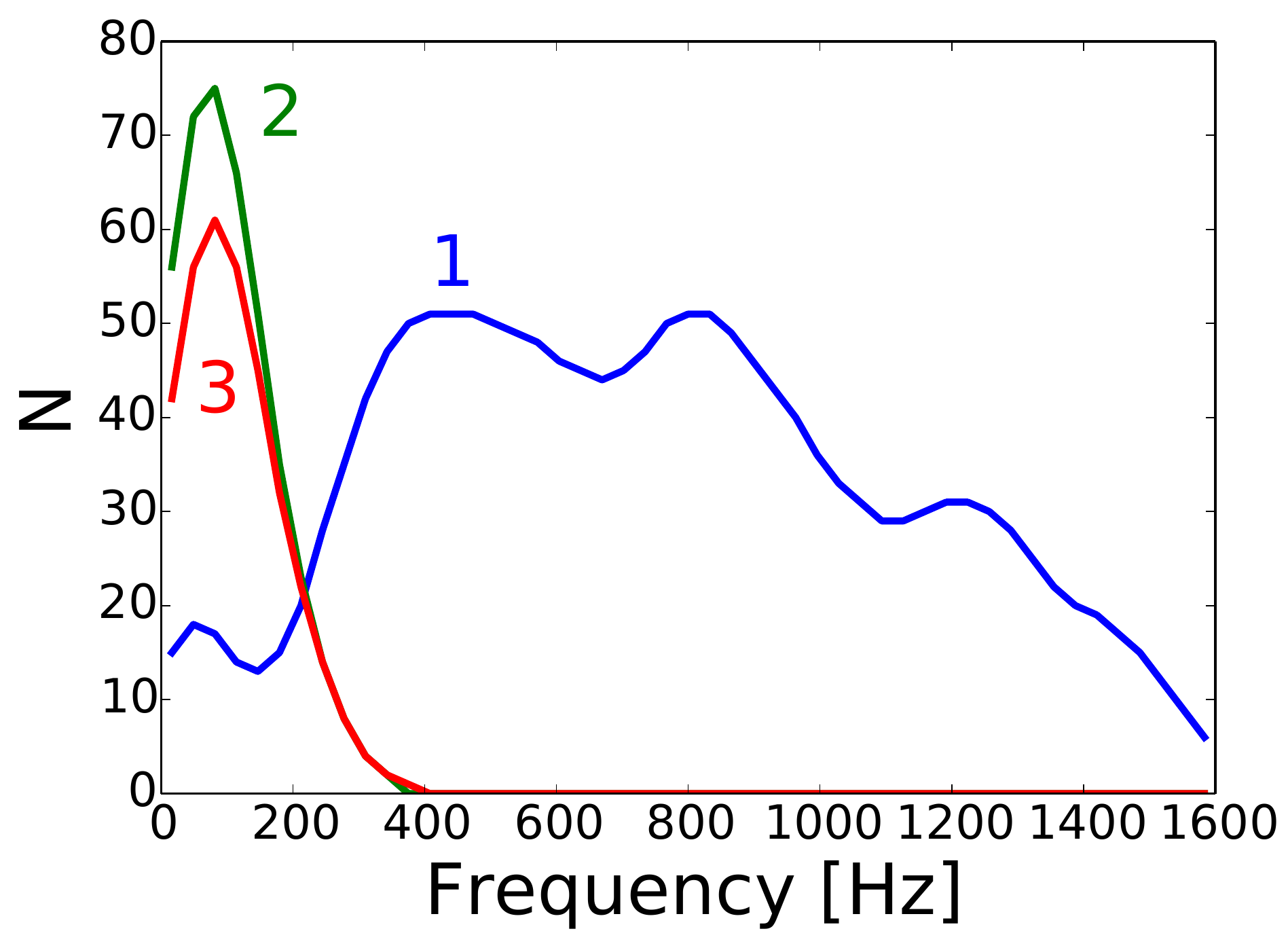}}
	\label{fig:MDC1-frequency-distribution}
\caption{The figure shows the reconstructed frequency distributions for the third data set for (a) PCAT (33 PCs), (b) LIB, and (c) WDF-ML. The numbers correspond to the different identified transient types.}
\end{figure}

%%%%%%%%%%%%%%%%%%%%%%%%%%%%%%%%%%%%%%%%%%%%%%%%%%%%%%%

\subsubsection{WDF-ML.}

WDF detected 2547/3000 of the noise transients in data set 3, using a threshold SNR value of 15. 
The sine Gaussian and Ring-down waveforms morphologies are mixed together in type 0. The Gaussian waveforms have been split between types 1 and 2. 
The frequency for each WDF type in the data set is shown in Figure 6(c). Type 2 contains the lower frequency Gaussian waveforms (up to $\sim 250$ Hz), and the type 1 Gaussian waveforms have frequencies as high as $\sim 1500$ Hz.
The Gaussian waveforms that were incorrectly classified into type 0 were Gaussian waveforms with a low SNR $(\sim 20)$. 

Choosing more components for the Spectral Embedding stage will result in more sub-types for the sine Gaussian and Ring-down waveforms, but no clear distinction between the two types. In this data set the noise transients are spread in frequency and duration, therefore, results could be improved by using a multi-window analysis. This is a feature that will be added to a future version of the WDF-ML algorithm.

%%%%%%%%%%%%%%%%%%%%%%%%%%%%%%%%%%%%%%%%%%%%%%%%%%%%

\subsubsection{Summary.}

All three methods were able to correctly classify the Gaussian noise transients included in this data set, but were unable to distinguish between the sine Gaussian and Ring-down waveform morphologies when the range of parameters for the waveforms was very large. This is because a low frequency sine Gaussian waveform has a closer waveform shape to a low frequency Ring-down waveform than to a high frequency sine Gaussian waveform. Real glitches with characteristic waveforms usually have narrow frequency or duration distributions, but this data set allows us to test the limitations of the different transient classifying algorithms. 

The wide range of parameters of the simulated waveforms, especially duration, make it difficult to capture the variability of the waveforms in the first few PCs. The PCAT and LIB classification could be improved by being more selective about which waveforms are included in the making of the PCs. WDF-ML may see similar improvements by altering the windowing parameter used in the analysis.

%%%%%%%%%%%%%%%%%%%%%%%%%%%%%%%%%%%%%%%%%%%%%%%%%%%%%%%%%%%%%%%%%%%%%%%%%%%%%%%%%%%%%%
%%%%%%%%%%%%%%%%%%%%%%%%%%%%%%%%%%%%%%%%%%%%%%%%%%%%%%%%%%%%%%%%%%%%%%%%%%%%%%%%%%%%%%
%%%%%%%%%%%%%%%%%%%%%%%%%%%%%%%%%%%%%%%%%%%%%%%%%%%%%%%%%%%%%%%%%%%%%%%%%%%%%%%%%%%%%%

\section{Discussion}
\label{discussion}

This paper introduces three new methods for the fast classification of noise transients in advanced gravitational-wave detectors,
and shows the results of testing and comparing these methods on data sets containing simulated noise transients. The purpose of 
this is to provide information that can lead to an improvement in data quality during a science run.   

We show that all three methods can classify transient waveforms in gravitational-wave detectors with a high level of efficiency. In our
first data set, which has transients well separated in frequency and SNR, over 97\% efficiency is obtained by all three methods. Reducing the threshold of the trigger generators, and therefore
including transients with an SNR less than 20, can reduce the classification efficiency. In the second data set we show that all 
three methods can classify noise transients by waveform morphology alone. This can be split into further types by frequency and
SNR if the number of types requested is bigger than the number of morphology types in the detector data. The third data set was more challenging
to classify due to the large range of parameters of the simulated transients. 

The different algorithms identified different numbers of signals in the data. To identify transients PCAT's trigger generator measures the excess power in the time series of a given channel. More sophisticated methods for transient identification have been devised, and they are in use in the LIGO and Virgo data analysis and detector characterization groups. However, the main goal of PCAT's algorithm is to provide a proof of concept for classification of transients rather than to provide a trigger generator for detector characterization analysis. Thus a simple identification method based on excess power in time bins is sufficient for our scope. Future plans for the use of the PCA technique include improving the trigger generator or to interface the PCAT code with an existing trigger generator already in use by the LIGO and Virgo Collaborations.

For PCAT and LIB the number of PCs that are used can have a large effect on the results of the classification. 
If too many PCs are used then an incorrect classification is given, due to some of the PCs consisting
of only noise. As we cannot eliminate the background noise from the glitch waveforms that are used to make the PCs,
we have found the best method of choosing the number of PCs to be the position of the ``knee" of the variance plots.    
For WDF-ML the selection of the analyzing window size for the wavelet transform is fundamental for a correct classification. 
The window must be larger than the length of the transients in the data, and to avoid a false classification of a noise transient the waveform must not be overlapping between two windows.

In this study we only use the gravitational-wave channel of the detector. As all signals in the data will be classified into glitch types it is possible that a real gravitational-wave signal could be included in our glitch classification results. This could be avoided by removing signals that are coincident between two detectors before applying the classification methods. In future work we plan to include multiple auxiliary channels
in the classification procedure. If a noise transient occurs in the gravitational-wave channel in time coincidence with an auxiliary 
channel, it can help us to identify the cause of the transient type \cite{2013PhRvD..88f2003B, 2013CQGra..30o5010E}. The number of possible auxiliary channels may be very large, which 
makes machine learning an ideal tool for this type of classification due to the speed at which machine learning methods can process a large
volume of detector data. 

PCAT runs daily on data from the aLIGO detectors, providing a powerful diagnostic tool to the Detector Characterization
team in preparation for the first aLIGO observing run (O1). LIB plans to start running daily on aLIGO data before the start of O1, and to provide information back to the Detector Characterization teams to be used during data quality shifts. WDF has been used as a noise transient event trigger generator, and monitoring tool, during past Virgo science runs. The machine learning classification procedure of WDF-ML is an innovative addition to this algorithm that will be used to classify transients during the advanced detector science runs. The algorithms can be run on parallel computing clusters, and the code can be optimised, to allow the algorithms to run efficiently in real time.

%%%%%%%%%%%%%%%%%%%%%%%%%%%%%%%%%%%%%%%%%%%%%%%%%%%%%%%%%%%%%%%%%%%%%%%%%%%%%%%
%%%%%%%%%%%%%%%%%%%%%%%%%%%%%%%%%%%%%%%%%%%%%%%%%%%%%%%%%%%%%%%%%%%%%%%%%%%%%%%
\ack

The authors thank Salvatore Vitale and Erik Katsavounidis for comments and advice regarding LALInference Burst. 
We thank the Burst and DetChar groups of the LIGO Scientific Collaboration for helpful
discussions of this work. 
DT and MC are partially supported by the National Science Foundation
through awards PHY-1067985 and PHY-1404139.
ISH and JP gratefully acknowledge the support of the
UK Science and Technology Facilities Council (STFC). 
JP, ISH and EC also gratefully acknowledge the support of the Scottish Universities Physics Alliance (SUPA).

%%%%%%%%%%%%%%%%%%%%%%%%%%%%%%%%%%%%%%%%%%%%%%%%%%%%%%%%%%%%%%%%%%%%%%%%%%%%%%%%
%%%%%%%%%%%%%%%%%%%%%%%%%%%%%%%%%%%%%%%%%%%%%%%%%%%%%%%%%%%%%%%%%%%%%%%%%%%%%%%%
%%%%%%%%%%%%%%%%%%%%%%%%%%%%%%%%%%%%%%%%%%%%%%%%%%%%%%%%%%%%%%%%%%%%%%%%%%%%%%%%

\section*{References}
\bibliographystyle{iopart-num}
\bibliography{bibfile}

%%%%%%%%%%%%%%%%%%%%%%%%%%%%%%%%%%%%%%%%%%%%%%%%%%%%%%%%%%%%%%%%%%%%%%%%%%%%%%%
\newpage
\appendix
\label{app:a}
\section{}

\begin{table}[h!]
\begin{centering}
\begin{tabular}{| l |  c  |  c  | c | }
  \hline
& Waveform & Min Value & Max Value \\ \hline
Frequency (Hz) & SG & 380 & 420 \\ 
hrss (${\rm Hz}^{-1/2}$) & SG & $1\times 10^{-21}$ & $5\times 10^{-21}$ \\ 
Q & SG & 5 & 10 \\ 
SNR & SG & 1 & 400 \\ 
hrss (${\rm Hz}^{-1/2}$) & G & $1\times 10^{-21}$ & $5\times 10^{-21}$ \\ 
Duration (s) & G & 0.001 & 0.01 \\ 
SNR & G & 1 & 400 \\ \hline
\end{tabular}
\label{SGtable}
\caption{The minimum and maximum parameters used when creating the simulated noise transients in data set 1.}
\end{centering}
\end{table}

\begin{table}[h!]
\begin{centering}
\begin{tabular}{| l |  c  |  c  | }
  \hline
   & Min Value & Max Value \\ \hline
   Frequency (Hz) & 40 & 1500 \\ 
   hrss (${\rm Hz}^{-1/2}$) & $5\times 10^{-22}$ & $4\times 10^{-21}$ \\ 
   Q (SG, RD) & 2 & 20 \\ 
   duration (G) & 0.001 & 0.01 \\ \hline
\end{tabular}
\label{Data1}
\caption{The minimum and maximum parameters used when creating the simulated noise transients in data set 3.} 
\end{centering}
\end{table}

\end{document}